\RequirePackage{fix-cm}
\documentclass[natbib,smallextended]{svjour3}       
\smartqed  
\makeatletter
\def\cl@chapter{\@elt {theorem}}
\makeatother
\usepackage[usenames,dvipsnames]{xcolor}
\usepackage[breaklinks,colorlinks,linkcolor=Maroon,citecolor=Blue]{hyperref}
\usepackage{amsmath}
\usepackage{amssymb}
\usepackage[utf8]{inputenc}
\usepackage[T1]{fontenc}
\usepackage[english]{babel}
\usepackage{graphicx}
\usepackage[retain-unity-mantissa = false]{siunitx}
\usepackage{xspace}
\usepackage[all]{hypcap} 
\usepackage{natbib}
\usepackage[colorinlistoftodos,textsize=tiny]{todonotes} 
\usepackage[capitalise,nameinlink]{cleveref}
\crefformat{equation}{#2Eq.~#1#3}
\Crefformat{equation}{#2Eq.~#1#3}
\newcommand{\mum}{\si{\um}}
\newcommand{\submm}{\mbox{(sub-)milli}\-meter\xspace}

\newcommand{\Hg}{\ensuremath{H_\mathrm{g}}\xspace}

\newcommand{\kb}{\ensuremath{k_\mathrm{b}}\xspace}
\newcommand{\cs}{\ensuremath{c_\mathrm{s}}\xspace}
\newcommand{\St}{\ensuremath{\mathrm{St}}\xspace}
\newcommand{\Vk}{\ensuremath{V_\mathrm{k}}\xspace}
\newcommand{\alphamm}{\ensuremath{\alpha_\mathrm{mm}}\xspace}
\newcommand{\rhos}{\ensuremath{\rho_\mathrm{s}}\xspace}
\newcommand{\tstop}{\ensuremath{t_\mathrm{stop}}\xspace}
\newcommand{\rhog}{\ensuremath{\rho_\mathrm{g}}\xspace}
\newcommand{\rhod}{\ensuremath{\rho_\mathrm{d}}\xspace}
\newcommand{\Sigg}{\ensuremath{\Sigma_\mathrm{g}}\xspace}
\newcommand{\Sigd}{\ensuremath{\Sigma_\mathrm{d}}\xspace}

 \journalname{my journal}
\begin{document}

\title{Dust Evolution and the Formation of Planetesimals
}


\author{T. Birnstiel \and
        M. Fang \and
        A. Johansen
}

\authorrunning{Birnstiel, Fang, Johansen} 

\institute{T. Birnstiel \at
            Harvard-Smithsonian Center for Astrophysics\\
            60 Garden Street, 02138 Cambridge, MA, USA\\
            \email{tbirnstiel@cfa.harvard.edu}\\
            Present address: Max Planck Institute for Astronomy, K\"onigstuhl 17, D-69117 Heidelberg
            \and
            M. Fang \at
            Purple Mountain Observatory and Key Laboratory for Radio Astronomy\\
            2 West Beijing Road, 210008,
            Nanjing, China\\
            \email{mfang@pmo.ac.cn}
            \and
            A. Johansen \at
            Lund Observatory, Department of Astronomy and
            Theoretical Physics, Lund University\\
            Box 43, 22100 Lund, Sweden\\
            \email{anders@astro.lu.se}
}

\date{Received: 6 August 2015 / Accepted: 11 April 2016}

\maketitle


\begin{abstract}
The solid content of circumstellar disks is inherited from the interstellar medium: dust particles of at most a micrometer in size. Protoplanetary disks are the environment where these dust grains need to grow at least 13 orders of magnitude in size. Our understanding of this growth process is far from complete, with different physics seemingly posing obstacles to this growth at various stages. Yet, the ubiquity of planets in our galaxy suggests that planet formation is a robust mechanism. This chapter focuses on the earliest stages of planet formation, the growth of small dust grains towards the gravitationally bound ``planetesimals'', the building blocks of planets. We will introduce some of the key physics involved in the growth processes and discuss how they are expected to shape the global behavior of the solid content of disks. We will consider possible pathways towards the formation of larger bodies and conclude by reviewing some of the recent observational advances in the field.
\end{abstract}

\clearpage

\section{Introduction}\label{sec:introduction}
Circumstellar disks consist mainly (99\% by mass) of gas, but the tiny 1\% of condensible material (commonly called solids or dust) is nevertheless a crucial ingredient. After all it is the material out of which planets and minor bodies are formed, but beyond this it is also important for the physical structure and the evolution of the disk: in most parts of the disk, the opacity is dominated by the dust. The dust thus determines the temperature structure of the disk by absorbing the stellar irradiation and reradiating it in the infrared. Consequently, dust determines the observational appearance of the disk: on the one hand through its thermal continuum emission and on the other hand by determining the temperature and density structure and therefore the excitation conditions for gas lines. Furthermore, solids provide the surface area for crucial surface chemical reactions (such as the formation of complex organics) and they influence the ionization levels in the disk by sweeping up free electrons. Finally, dust is also a key observational probe of the gas dynamics as it reacts sensitively to changes in the gas disk (see \cref{sec:dust-dynamics}). Clearly, understanding the evolution of solids is a key element in the puzzle of planet formation. 

The evolution of solids in a circumstellar disk is governed by transport processes and by collisional processes. Both categories will be discussed in the following, but it is important to note that both effects strongly depend on each other: transport processes typically depend on the particle size, hence the collisional evolution of the particles, while in turn, the collisions between the particles are driven by the dynamics. Studying transport or particle growth separately may be a useful exercise, but for a self-consistent picture of the global evolution of solids, both processes need to be considered together.

Given the broadness of this review, we can only scratch the surface of these topics. For more in-depth reviews the reader may refer to the relevant reviews in Protostars and Planets VI \citep[for example,][]{Johansen:2014hs,Testi:2014cj,Espaillat:2014hh,Turner:2014ee}, to the reviews by \cite{Armitage:2011fm}, \citet{Williams:2011js}, and \citet{Andrews:2015cg}. In the following we will introduce first some of the basic transport mechanisms (\cref{sec:dust-dynamics}) before discussing the collisional evolution of dust (\cref{sec:dust-growth}). We will then put both pieces together to understand the global distribution of dust in the disk (\cref{sec:dust-disk}) and how planetesimals, the building blocks of planets can be formed (\cref{sec:planetesimal-formation}). Finally, \cref{sec:observations} will review how recent observations can help us constrain the evolution of solids in circumstellar disks and \cref{sec:summary} summarizes this chapter.

\section{Dust Dynamics}
\label{sec:dust-dynamics}

\subsection{Drag Forces}
\label{sec:dust-dynamics:drag-forces}
Aerodynamic drag is a phenomenon known from daily life, may it be the head wind experienced on a bicycle or the tail wind that accelerates a sailing boat: whenever there is a difference in velocity between an object and the surrounding gas, the drag force acts towards eliminating the velocity difference. The same principles apply to solids in circumstellar disks that are dynamically coupled to the gas via drag forces. However in most regions of the disk, the drag stems not from the flow around the object as we experience it. Instead, most dust grains are smaller than the mean free path of the gas molecules and they rather feel drag by being bombarded by gas molecules slightly faster from the direction of the flow than behind it, effectively a pressure force. This drag force is called Epstein drag (in contrast to the better known Stokes drag, which can become relevant in the denser inner regions of disks). The Epstein drag force is expressed as \citep{Weidenschilling:1977p865}
\begin{equation}
\vec F_\mathrm{Ep} = - \frac{4\pi}{3}\rhog \,a^2 \, \Delta\vec{\mathit{w}} \, v_\mathrm{th},
\end{equation}
where \rhog is the gas density, $a$ the particle radius, $\Delta\vec{\mathit{w}}$ the particle-gas relative velocity, $v_\mathrm{th} = \sqrt{8/\pi}\cs$ the mean thermal velocity, and \cs the isothermal sound speed. Instead of using this definition of the force, it is much more useful to specify the \textit{stopping time}
\begin{equation}
\tstop = \frac{m\,\Delta w}{F_\mathrm{Ep}} = \frac{\rhos\,a}{\rhog\,
v_\mathrm{th}}
\label{eq:stopping-time}
\end{equation}
where we assumed that mass and radius are related as $m=4\pi\,\rhos\,a^3 /3$ with a mean material density of the dust particle \rhos. For fractal particles, the mass is still well defined, but their size or cross section can be defined in various ways \citep[e.g.][]{Ormel:2007bh,Okuzumi:2009hf} and the mass-size relation follows a different power-law than 3 (called the fractal dimension). To give an example, for typical disk mid-plane conditions at \SI{1}{AU} (e.g. $\Sigg=$ \SI{200}{g.cm^{-2}}), the stopping time is about a few seconds for a compact micrometer sized particle and about 10 days for a decimeter sized particle. Even more useful than the stopping time is the dimensionless ratio of the stopping to the dynamical time scale, called \textit{Stokes number},
\begin{equation}
\St = \tstop\,\Omega,
\label{eq:stokes-number}
\end{equation}
where $\Omega$ is the Keplerian angular velocity. The Stokes number is so useful because two particles of different composition, structure, mass, \ldots behave aerodynamically identical if their Stokes numbers are identical. For a vertically isothermal gas disk with scale height $\Hg=\cs/\Omega$ and gas surface density $\Sigg$, the Stokes number at the disk mid-plane becomes
\begin{equation}
\St = \frac{a\,\rhos}{\Sigg}\frac{\pi}{2},
\end{equation}
where we have assumed a gas mid-plane density of $\rho_\mathrm{g,mid} = \frac{\Sigg}{\sqrt{2\pi}\Hg}$.
Under these assumptions, the Stokes number is linearly dependent on the particle size, which is true for most but the densest regions of the disk. In the following we will mostly talk about Stokes numbers instead of particle sizes, but the Stokes number can just be viewed as a dimensionless quantity describing the particle size: a small particle with a small Stokes number ($\St\ll 1$) will be adapting to the gas velocity on time scales much shorter than the orbital time scale, while a big particle with very large Stokes number ($\St\gg1$) will perform several orbits before the drag forces significantly alter its velocity.

From this simple concept of size dependent drag forces, a surprising number of ``complications'' arise: the velocities of particles in a protoplanetary disk generally become size dependent (unlike the velocity of a dust grain in a free Keplerian orbit). Thus, the trajectory of a particle depends on its size. Different-sized particles having different velocities means that particles collide with each other and those collisions can lead to sticking/growth or to shattering/destruction of the particles. Other effects caused by the drag are: a systematic drift motion of particles (see next section), turbulent mixing of dust particles (\cref{sec:dust-dynamics:turbulent-mixing}), and dynamical instabilities caused by the coupling of dust and gas (\cref{sec:planetesimal-formation}).

\subsection{Dust Drift}
\label{sec:dust-dynamics:dust-drift}
It was already found by \citet{Whipple:1972vv} and \citet{Weidenschilling:1977p865} that dust particles embedded in a gaseous disk should migrate towards the star on short time scales. A derivation of the dust and gas velocities can also be found in \citet{Nakagawa:1986cu}. The velocities of dust ($\mathbf{u}_\mathrm{d}$) and of the gas ($\mathbf{u}_\mathrm{g}$) evolve according to two equations
\begin{align}
\frac{\mathrm{d}\mathbf{u}_\mathrm{d}}{\mathrm{d}t} &= - \frac{1}{\tstop} (\mathbf{u}_\mathrm{d}-\mathbf{u}_\mathrm{g}) -  \frac{\mathrm{G} M_\star}{r^3} \,\mathbf{r},\\
\frac{\mathrm{d}\mathbf{u}_\mathrm{g}}{\mathrm{d}t} &= - \frac{\epsilon}{\tstop} (\mathbf{u}_\mathrm{g}-\mathbf{u}_\mathrm{d}) -  \frac{\mathrm{G} M_\star}{r^3} \,\mathbf{r} - \frac{\nabla P}{\rho_\mathrm{g}},
\end{align}
where $\epsilon$ is the dust-to-gas density ratio ($\rho_\mathrm{d}/\rho_\mathrm{g}$), $P$ the gas pressure, $\mathrm{G}$ the gravitational constant, and $M_\star$ the stellar mass. Pressure acceleration on the dust particles is negligible (as the dust material density\footnote{It should be noted that \rhod denotes the total mass of dust particles per ``volume of space'', while \rhos denotes the specific weight of a dust particle, in other words the mass per ``dust volume''.} is much smaller than the gas density$\rhos\ll\rho_\mathrm{g}$). Rewriting these equations in cylindrical coordinates, assuming a steady state ($d/dt = 0$), a low dust-to-gas ratio, and then solving for the first order deviation from the Keplerian velocity ($\mathbf{v} = \mathbf{u}-\mathbf{\Vk}$) leads to the dust drift speed
\begin{align}
v_r		&\simeq - \frac{2}{\St+\St^{-1}} \,\eta\, \Vk\label{eq:radial_drift},\\
v_\phi	&\simeq - \frac{1}{1+\St^2}\, \eta\, \Vk,
\end{align}
where
\begin{equation}
\eta = -\frac{1}{2} \left(\frac{H_\mathrm{p}}{r}\right)^2 \frac{\partial\ln P}{\partial \ln r}
\end{equation}
describes how much slower than Keplerian the gas is orbiting, i.e. $v_{\phi,\mathrm{g}} = \eta\, \Vk$.
Our assumption of low dust-to-gas ratios basically means that the dust feels the drag force by the gas, but the effect on the gas velocity is negligible. The velocities for dust and gas for arbitrary dust-to-gas ratios can be found in \citet{Nakagawa:1986cu}. These results have several important consequences:

\begin{itemize}
  \item Particles with a Stokes number $< 1$ drift inward with a speed of $v_r \simeq -2 \,\St\,\eta\,\Vk$: so small particles move slowly, larger particles move faster. 
  \item For typical disk conditions, $\eta$ is of the order of a few per mille, which means that the maximum radial drift velocity for $\mathrm{St}=1$ is a few per mille of the Keplerian speed. In other words, the orbit of a dust particle decays on a time scale of only a few hundred orbits.
  \item The azimuthal drift velocity for particles with a Stokes number $<1$ is $\eta\,\Vk$, so they move along with the gas while the drift speed of larger particles tends towards zero, they move on Keplerian orbits.
  \item The direction of the radial drift is towards higher pressure, so generally inwards in a disk that is denser and hotter closer to the star.
\end{itemize}

The last point also applies to the vertical dimension: a particle at a height $z$ above the disk mid-plane would orbit on an inclined Keplerian orbit, so effectively oscillate around the mid-plane. The gas disk however will decelerate this motion via gas drag. Let us assume that the gas disk is vertically stable, so $u_{z,\mathrm{gas}}=0$. A dust particle with a Stokes number \St will by definition decelerate on time scales of \St $\times$ the orbital time scale. Particles with $\St>1$ would therefore execute a damped oscillation, while particles with a small Stokes number quickly reach a terminal settling velocity when the vertical acceleration due to the stellar gravity ($\dot u = - \Omega^2\,z$) and the deceleration from the drag force ($\dot u = u/\tstop$) are in balance. This is the velocity at which particles sediment towards the mid-plane,
\begin{equation}
u_{z,\mathrm{dust}} = - z\,\Omega\,\St.
\label{eq:v_settling}
\end{equation}

Finally, it should be noted, that particles also move azimuthally towards higher pressure. Even slight azimuthal over densities in the gas are thus able to produce very strong dust over-densities \citep{Birnstiel:2013es}. For example vortex structures are azimuthal over-densities that effectively trap dust particles \citep[e.g.,][]{Barge:1995vd,Klahr:1997cp,Lyra:2013ix,Raettig:2015vc}. Eccentric gas disks are the exception to the rule: azimuthal over-densities in eccentric disks stem from velocity variations along the eccentric orbit. These velocity (and thus density-) modulations are identical for dust and gas and therefore do not cause strong accumulations of dust particles \citep{Hsieh:2012en,Ataiee:2013kx}.

In addition to this drift motion, dust is also carried along with the radial gas flow. \citet{Takeuchi:2002jf} derived this component of the dust radial velocity to be
\begin{equation}
u_{r,\mathrm{dust}} = \frac{1}{1+\St^2}\, u_{r,\mathrm{gas}}.
\end{equation}
Hence small particles ($\St<1$) follow the gas flow, while large particles ($\St\gg\,1$) are unaffected by it.

The ideas presented in this section are based purely on theoretical concepts. Until recently, there has been very little observational evidence in support of radial or azimuthal segregation of dust particles (but some on vertical settling). Recently, this situation has changed dramatically and we will discuss the current observational support of these concepts in \cref{sec:observations}.

\subsection{Turbulent Mixing}
\label{sec:dust-dynamics:turbulent-mixing}

It is widely believed that turbulent effective viscosity is the driver of disk evolution \citep[this picture has recently been put into question by models where the angular momentum is transported by disc winds and the gas motion remains laminar, see][and references therein]{Turner:2014ee}. If the gas is indeed turbulent, then the dust motion and transport will be affected by it. In the following, we will assume, that the turbulence is described by an effective viscosity \citep{Shakura:1973uy}
\begin{equation}
\nu = \alpha\,\frac{\cs^2}{\Omega},
\label{eq:nu_gas}
\end{equation} 
where $\alpha$ is the turbulent strength parameter. The random motion induced by this turbulence will act as a diffusivity on the dust, and again, this will depend on the dust particle size. One can imagine that large boulders might be less effectively mixed than micrometer sized dust particles. The ratio of the dust diffusivity to the gas diffusivity is commonly called the Schmidt number\footnote{There are several definitions interchanging dust/gas or  diffusivity/viscosity. Here we follow the definition of \citet{Youdin:2007ef} who also discusses the different meanings of these notations.} $\mathrm{Sc}$ which was shown to be \citep{Youdin:2007ef} 
\begin{equation}
\mathrm{Sc} = \frac{D_\mathrm{g}}{D_\mathrm{d}} \simeq 1+\St^2,
\end{equation}
and it is commonly assumed that the gas diffusivity $D_\mathrm{g}$ equals the gas viscosity $\nu$ \citep[but see][]{Johansen:2005eq,Fromang:2006hk,Pavlyuchenkov:2007gw}. Turbulent mixing will result in smoothing out concentrations of trace species (dust grains, or molecular species). Such concentrations might be caused by dust drift (in azimuthal, vertical, or radial direction) or by local production. It might also be responsible for mixing thermally processed particles from the hot inner regions throughout the disk \citep[e.g.,][]{BockeleeMorvan:2002jx,Pavlyuchenkov:2007gw}. We will discuss some of these effects shaping the global appearance of the disk in the following section. Clearly this review can introduce only some of the key concepts. Several other effects, such as radiation pressure \citep[e.g.,][]{Dominik:2011ih} or photophoresis \citep[e.g.,][]{Krauss:2005gq} can be relevant in disks as well, but are not discussed here.

\section{Dust Growth}
\label{sec:dust-growth}
The initial stages of planet formation in the core accretion scenario necessarily involve the growth from sub-micrometer sized dust grains to $>$km sized bodies, called planetesimals, that are gravitationally bound. To understand how particles grow along so many orders of magnitude in mass, we need to understand (1) at which velocities particles collide, (2) how often particles collide, and (3) what the outcome of each collision is. The derivation of the collisional rates is straight forward: a particle $i$ sweeping through a swarm of particles $j$ has a mean free path of
\begin{equation}
l = \frac{1}{n_j\,\sigma_{ij}},
\end{equation}
where $n_j$ is the number density of particles $j$ and $\sigma_{ij}$ is the cross section of particles $i$ and $j$. Assuming spherical radii of $a_i$ and $a_j$ respectively, $\sigma_{ij} = \pi\,(a_i + a_j)^2$. If particles $i$ and $j$ move with a relative velocity of $\Delta v_{ij}$, this means, a particle $i$ will on average feel one collision per collisional time scale
\begin{equation}
\tau_\mathrm{col} = \frac{l}{\Delta v_{ij}} = \frac{1}{n_j\,\sigma_{ij}\,\Delta v_{ij}}.
\label{eq:t_col}
\end{equation}
If there are $n_i$ particles per volume, each feeling one collision with particles $j$ per collisional time scale $\tau_\mathrm{col}$, this means
\begin{equation}
R_{ij} = n_i\,n_j\,\sigma_{ij}\,\Delta v_{ij}.
\label{eq:coll_rates}
\end{equation}
This is the rate at which particles $i$ collide with particles $j$ (e.g., forming a new species $k$). This shows that the particle-particle velocities determine not only the outcome of a collision, but also how often it happens and that the rates depend on density squared and the cross section, both of which are outcomes of the growth process itself. Solving for the evolution of a particle size distribution means that we need to calculate the rates for all the collisions of all particle sizes and keep book where the results of a collision end up. As an example, let us consider pure sticking: this means that a collision of two particles with masses $m_i$ and $m_j$ will produce one particle with mass $m_k = m_i+m_j$ (a gain term for $n_k$), while every collision of a particle $m_k$ will lead to a loss term, hence
\begin{equation}
\dot n_k = \frac{1}{2} \sum_{i,j} R_{ij} \delta_{i,j-k} - \sum_i\,R_{i,k}, 
\end{equation}
where the factor of 1/2 comes from double counting collisions. In practice, the analytical or numerical solutions of grain growth are complicated by other collisional outcomes (fragmentation, cratering, \ldots) and by the fact that the grid spacing of the mass dimension cannot be linear. For example growing a 1~cm sized aggregate out of micrometer sized constituents (monomers) would require a grid with $10^{12}$ entries, which is computationally unfeasible. Logarithmic grid spacing however comes at the cost of having particles grow to masses where there is no grid point. The mass needs to be distributed over neighboring grid points and the algorithm becomes less accurate. Monte-Carlo methods with discrete particles can overcome these issues, but they suffer from other problems, such as time step constraints \citep[see, ][and references within]{Drazkowska:2014hp}.

\subsection{Impact Velocities}
\label{sec:dust-growth:impact-velocities}
If the velocity of a particle depends on its size (as is the case for radial, azimuthal, and vertical drift motion), particles of different sizes will have a relative velocity with respect to each other. For the radial, azimuthal, and vertical velocities above, the relative speed between two particles with Stokes numbers $\St_i$ and $\St_j$ becomes
\begin{align}
\Delta v_{r}(i,j)    &= \eta\,\Vk \, \left|\frac{2}{\St_i+\St_i^{-1}} - \frac{2}{\St_j+\St_j^{-1}} \right|\\
\Delta v_{\phi}(i,j) &= \eta\,\Vk \,\left| \frac{1}{1+\St_i^2} - \frac{1}{1+\St_j^2}\right|\\
\Delta v_{z}(i,j)    &= z\,\Omega\,\left|\St_i-\St_j \right|.
\end{align} 
As we can see, for $\St \ll 1$, the azimuthal velocities vanish, and the radial velocities become proportional to $\left|\St_i-\St_j\right|$. We also note that for $\St_i=\St_j$, all velocities vanish, as particles move with the same systematic velocities.

The latter is not the case for random motions. In protoplanetary disks, two effects cause random motion of dust particles: Brownian motion and turbulence. In the case of Brownian motion, the mean thermal kinetic energy $E_\mathrm{kin}= \frac{1}{2} m v_\mathrm{th}^2$ at a temperature $T$ is distributed equally amongst the particles. The average relative velocity of two particles with masses $m_i$ and $m_j$ with a Maxwell-Boltzmann distributed velocity becomes
\begin{equation}
v_\mathrm{BM} = \sqrt{\frac{8 \,k_\mathrm{B}\,T (m_i+m_j)}{\pi\,m_i\,m_j}},
\label{eq:v_brownian}
\end{equation} 
where $k_\mathrm{B}$ is the Boltzmann constant, $\mu$ the mean molecular weight and $m_\mathrm{p}$ the proton mass. In contrast to the relative velocities discussed above, this one does not vanish for equal mass particles.

The case of turbulent velocities is much more complicated: particle velocities are affected by turbulent eddies through which they move. Turbulent eddies of different sizes cause partial alignment of particle trajectories on short distances or induce random kicks, depending on the eddy turnover time, the particle stopping time, and the time it takes the particle to cross the eddy. The classical picture was introduced by \citet{Volk:1980vu} and allowed \citet{Ormel:2007bl} to derive closed form expressions that are currently widely used. \citet{Pan:2010hi}, \citet{Pan:2014ii} and following papers in this series criticized this picture, emphasizing that the importance of the particle separation prior to a collision in determining the degree of correlation between particle velocities induced by turbulent eddies of different sizes. Their results deviate somewhat from the results of \citet{Ormel:2007bl} (predicting a maximum velocity by about a factor of two lower) and they additionally allow the calculation of the distribution of collision velocities. In the following examples, we will use approximations to the velocities derived by \citet{Ormel:2007bl} for simplicity. In this framework, the collision speed depends on the particle sizes of both particles and on the largest and smallest eddy turn-over times. For equal sized particles with stopping times larger than the smallest eddy turn-over times, the velocities can be approximated by
\begin{equation}
\Delta v_\mathrm{turb} \simeq \sqrt{\frac{3\,\alpha}{\St+\St^{-1}}}\,\cs,
\label{eq:v_turbulent}
\end{equation}
where $c_\mathrm{s}$ is the isothermal sound speed. Similar to relative velocities from dust drift, also turbulent collision velocities increase with Stokes number ($\simeq$ particle size), then reach a maximum at $\St=1$ and decrease for particles with $\St>1$.

\begin{figure}[htp]
\begin{center}
  \includegraphics[width=\hsize]{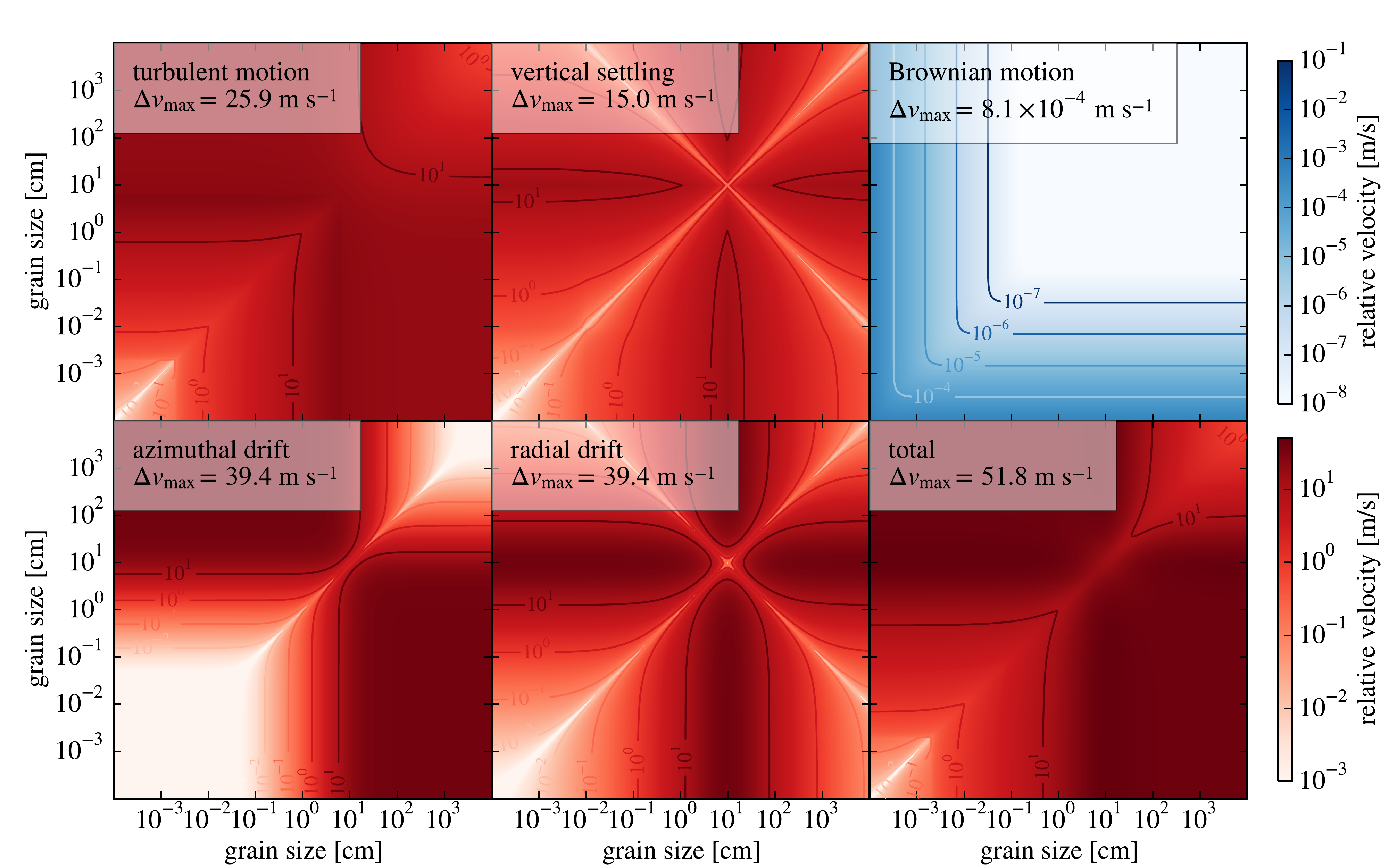}
  \caption{Various contributions to particle mean relative velocities in protoplanetary disks: turbulence, vertical settling, azimuthal and radial drift yield maximum impact speeds of tens of meters per second (bottom/red color scale). Brownian motion is the dominant source of relative velocities for small particles (top/blue color scale). The vertical settling velocities plotted here were calculated by using the settling velocity at the respective dust scale height.}
  \label{fig:rel_vel}
\end{center}
\end{figure}

Comparing the terms above, we can see that all velocity contributions vanish towards small particle size, apart from Brownian motion (see also \cref{fig:rel_vel}). Hence, initial growth is seeded by the Brownian velocities and only once particles have grown larger than a few micrometers, turbulent relative velocities start to become important \citep{Birnstiel:2011ks}. At larger sizes, radial drift velocities are dominating\footnote{As discussed above, for equal sized particles, drift velocities vanish, but typically the particle size dispersion causes particles to collide with particles of similar sizes.}, unless turbulence is quite strong \citep[$\alpha\gtrsim 2 (H/r)^2$, see also][]{Testi:2014cj}. For particles  with $\St\gtrsim 1$, azimuthal relative velocity contributions are the strongest contribution before gravitational perturbations from the turbulent gas density field come into play \citep{Johansen:2014hs}.

The various contributions to relative particle velocities are plotted in \cref{fig:rel_vel} at 10~AU in a disk with $\alpha=10^{-3}$, a local temperature of \SI{63}{K} and a gas surface density of \SI{16}{g.cm^{-2}}. The maximum relative velocity from turbulent velocities is given by the gas root mean squared velocity $v_\mathrm{rms} = c_\mathrm{s}\,\sqrt{3\,\alpha/2}$ (see \cref{eq:v_turbulent}). These velocities are quite low compared for example to the interstellar medium (for a temperature of \SI{100}{K} and typical turbulence parameters of $\alpha = 10^{-3} \ldots 10^{-2}$ they range above 20 -- 70 \si{m.s^{-1}}), but they are still high enough to destroy large particles upon collision. Similarly, drift-induced relative velocities reach a few times $c_\mathrm{s}\,H/r$, which also works out to similar numbers ($\lesssim$~\SI{80}{m.s^{-1}}). Collision speeds of micrometer sized bodies, on the other hand, are very small, of the order of millimeters per second. These numbers already indicate that we can expect a wide variety of collisional outcomes and that the outcomes change significantly as particles grow and collide at larger and larger velocities.

\subsection{Collisional Outcomes}
\label{sec:dust-growth:collisional-outcomes}

The most important ingredient for models of particle growth are \textit{collisional outcome models}: given two particles and the collision speed, a collisional outcome model predicts the properties of the resulting particle(s). Realistically, this outcome depends not only on the impact velocity, but also on impact parameter, size of the grains, compositional properties of both grains such as porosity, monomer size, fractal dimension, chemical composition (surface ices, hence temperature), and other properties. Obviously this entire parameter space of particle collisions cannot possibly be explored comprehensively with laboratory studies. Still, many studies have shed light on this enormous parameter space and identified possible outcomes and how they depend on or scale with particle properties and collisional parameters \citep[see][for recent reviews]{Blum:2008fi,Testi:2014cj,Johansen:2014hs}. Other studies have condensed these results into collision models, using existing laboratory results and experimentally or theoretically constrained scaling relations. Examples include \citet{Suyama:2008bx}, \citet{Guttler:2010ia}, \citet{Zsom:2010hg}, \citet{Windmark:2012gi}, or \citet{Krijt:2015bu}. The main collisional outcomes are:
\begin{itemize}
  \item \textit{sticking:} hit-and-stick collisions
  \item \textit{bouncing:} particles bouncing off each other without changing their mass, possibly causing compaction
  \item \textit{erosion/cratering:} smaller projectile removes mass from larger target, possibly shattering itself
  \item \textit{mass transfer:} smaller projectile fragments upon collision with target, but also deposits some fraction of its mass
  \item \textit{fragmentation:} complete destruction of the particle(s), fragments are typically distributed in a power-law fragment size distribution 
\end{itemize}

\begin{figure}[htp]
\begin{center}
  \includegraphics[width=\hsize]{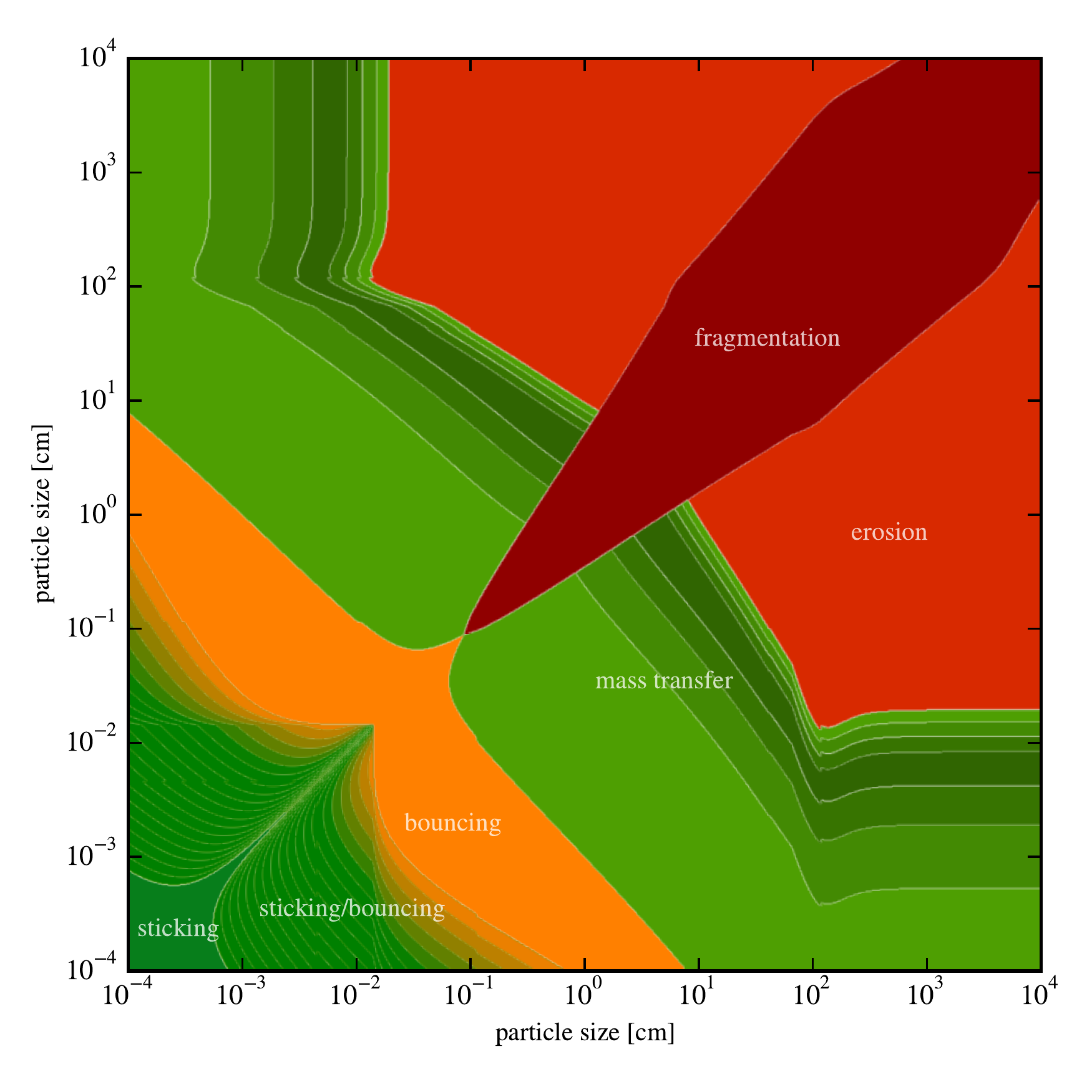}
  \caption{Mean collisional outcomes for silicate grains from \citet{Windmark:2012bg} as expected for a minimum-mass solar nebula disk \citep{Weidenschilling:1977kq} at 1~AU, including all the contributions of relative velocities discussed in \cref{sec:dust-growth:impact-velocities}. \textit{Green} regions denote net-growth of the larger collision partner, \textit{red} mass loss, and \textit{orange} denotes mass-neutral bouncing collisions.}
  \label{fig:windmark}
\end{center}
\end{figure}

\cref{fig:windmark} shows a typical result of the collisional model of \citet{Windmark:2012gi}, where, given two particle sizes, the \textit{mean} impact velocity is calculated (e.g. like in \cref{fig:rel_vel}), and a collisional outcome is predicted. A coagulation algorithm can thus evolve a size distribution of particles $n(m)$ by calculating the collision rates according to \cref{eq:coll_rates} and the collisional outcome model then determines how the rates are connected to the gain and loss terms, or in other words: how collisions between two particles $i$ and $j$ affect the number of particles $k$. Detailed descriptions of different recent astrophysical coagulation codes can be found in \citet{Brauer:2008bd}, \citet{Laibe:2008bc}, \citet{Ormel:2007bh}, \citet{Ormel:2008bp}, \cite{Zsom:2008je}, \citet{Okuzumi:2009hf}, \citet{Birnstiel:2010eq}, \citet{Okuzumi:2012kd}, and references within.

It should be highlighted, that, as mentioned in the beginning of this section, the collision outcome is not only a function of particle masses and velocities, but also other particle properties, and that the collision velocity of two particles is not a fixed mean velocity, but a distribution of velocities. The effect of the velocity distribution on the outcome can be substantial, as shown in \citet{Windmark:2012bg} and \citet{Garaud:2013ja}. Nevertheless, the mean outcome gives a good estimate for how the bulk of the particle distribution is evolving, as we will see in the following section.

\subsection{Simple estimates}
\label{sec:dust-growth:estimates}

Coagulation/fragmentation processes are not local in mass space. Numerically speaking, particles of one mass bin are not only interacting with neighboring cells (like in hydrodynamics). Instead each particle size can interact with every other particle size and influence the entire distribution, for example by producing a distribution of fragments. This fact makes a full treatment of the growth processes conceptually and numerically difficult. In many cases, however, growth proceeds in a more or less ordered fashion: small particles grow steadily and they grow most efficiently with particles of similar mass. Hence, simple estimates of particle growth are not only very instructive, they can also give reasonable answers, e.g. on how the upper end of the particle size distribution evolves. To derive such estimates, we will consider the case of \textit{monodisperse growth}, i.e. assuming that all particles are of one radius $a$. If particles only stick to each other, the mass of the newly formed particle is twice the original mass. The rate of collisions of a particle is then one collision per collisional time $\tau_\mathrm{col}$ (\cref{eq:t_col}). Upon each collision, the mass of the particle is doubled, so
\begin{equation}
\dot m \simeq \frac{m}{\tau_\mathrm{col}} = \rho_\mathrm{d} \,\sigma\,\Delta v.
\end{equation}
Now the cross section of two spherical particles of size $a$ is given by $\sigma = 4\,\pi\,a^2$ and we assume a constant porosity (i.e., $\rho_\mathrm{s}=const.$), so $\mathrm{d}m/\mathrm{d}a = 4\pi \rho_\mathrm{s}\,a^2$. This leads to the growth rate 
\begin{equation}
\dot a = \frac{\rho_\mathrm{d}}{\rho_\mathrm{s}}\,\Delta v.
\label{eq:da_dt}
\end{equation}
Now if the particle collision velocity is set by Brownian motion (\cref{eq:v_brownian}), then
\begin{equation}
\dot a = \underbrace{
	\frac{\rho_\mathrm{d}}{\pi}\sqrt{\frac{12\,k_\mathrm{B}\,T}{\rho_\mathrm{s}^3}}
}_{:=A}
\,a^{-3/2},
\end{equation}
which can be integrated to \citep[e.g.,][]{Dullemond:2005hf}
\begin{equation}
a(t) = \left[a(t_0)^{5/2}+\frac{5}{2}\,A\,t\right]^{2/5}.
\end{equation}
Similar calculations can be done for other relative velocities and they produce trends of numerical simulations well \citep[e.g.,][]{Ormel:2009dq,Birnstiel:2010eq,Windmark:2012gi}. These estimates obviously break down at sizes, where other effects come into play, for example when bouncing or fragmentation limits further growth.

\section{The Structure of the Dust Disk}
\label{sec:dust-disk}

In this section, we want to discuss how dust transport and growth shapes the global structure of the disk. For this we have to consider that particle transport depends on the particle size (which is evolving over time) while the particle size evolution in turn depends on local conditions (temperature, gas density, dust density, \ldots) and the relative velocities, which depend on the dynamical history of the particles. Thus particle growth and transport go hand in hand and to understand global trends, both effects need to be taken into account in a self-consistent way.

\subsection{Vertical Structure}
\label{sec:dust-disk:vertical}
\begin{figure}[htp]
\begin{center}
  \includegraphics[width=\hsize]{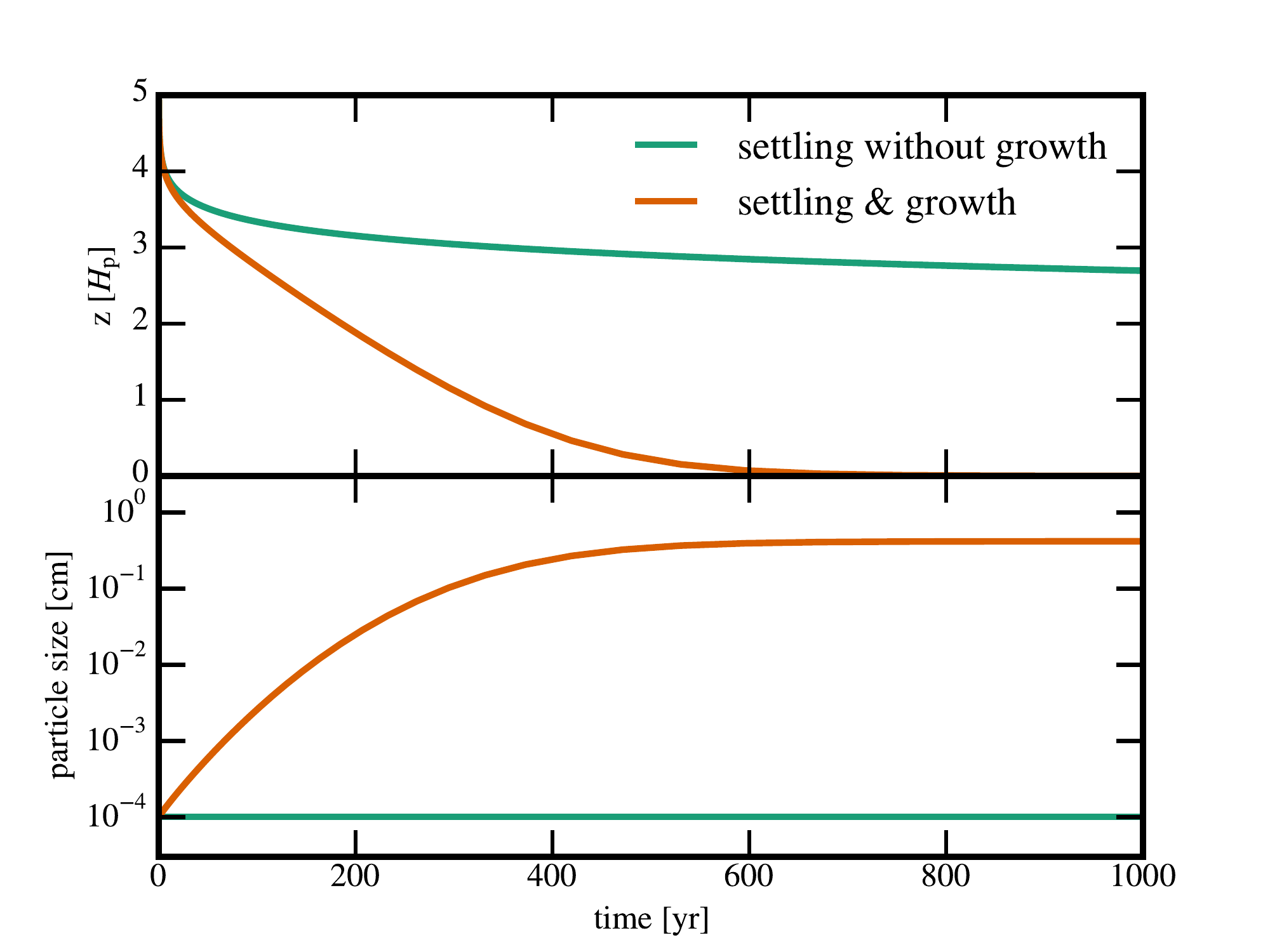}
  \caption{Trajectory of a settling particle at 1~AU. Figure and parameters after \citet{Dullemond:2005hf}. The initial particle size is one micrometer and the initial position is $5~\Hg$ above the mid-plane. The green curve integrates the trajectory assuming a constant particle size, the orange curve assumes sweep-up of other (assumed fixed) particles after \citet{Safronov:1969tr}.}
  \label{fig:settling}
\end{center}
\end{figure}

As explained in \cref{sec:dust-dynamics:dust-drift,sec:dust-dynamics:turbulent-mixing}, particles sediment towards the mid-plane but turbulence is counteracting this concentration effect. The settling time scale is $t_\mathrm{sett} = z/v_\mathrm{sett} = (\Omega\,\St)^{-1}$. Even small, micrometer sized particles have a large Stokes number if the density is as low as in the disk atmosphere, thus even the smallest particles sediment quickly. However, if they do not grow, their Stokes number will decrease as they settle towards higher gas densities and their downward motion will slow down (see green curves in \cref{fig:settling}). If they continue to grow while settling down, their increase in radius can partly counteract the increase in gas density and their Stokes number can stay large enough, such that they continue to settle towards the mid-plane, reaching macroscopic sizes along the way (see orange lines in \cref{fig:settling}). 

Turbulent mixing acts as diffusion on the dust distribution. We can estimate the combined effects of settling and mixing in the following way: the diffusion time scale over a vertical scale $z$ is $t_\mathrm{diff} = z^2/D_\mathrm{d}$. Setting this in relation to the settling time scale, we find
\begin{equation}
\mathrm{Pe} = \frac{t_\mathrm{diff}}{t_\mathrm{sett}} = \frac{\St}{\alpha}\, \left(\frac{z}{\Hg}\right)^2,
\end{equation}
where $\mathrm{Pe}$ is called the P{\'e}clet number. For large $z>\Hg\sqrt{\alpha/\St}$, the P{\'e}clet number is $>1$, meaning that settling time scales are shorter than diffusion time scales. Close to the mid-plane, $\mathrm{Pe}<1$ and diffusion is dominating. Thus, we can already expect the vertical dust-to-gas ratio to drop significantly above a dust scale height
\begin{equation}
H_\mathrm{d} = \Hg\,\sqrt{\frac{\alpha}{\St+\alpha}},
\label{eq:h_dust}
\end{equation}
similar to what was found by \citet{Dubrulle:1995jn}. The additional summand $\alpha$ stems from the fact that $H_\mathrm{g}$ is an upper limit for the dust scale height for the case of perfect mixing, i.e. $\alpha\gg\mathrm{St}$. More generally, the vertical structure and its evolution is described by an advection-diffusion equation \citep[e.g.][]{Dubrulle:1995jn,Schrapler:2004jj,Johansen:2005eq,Fromang:2009ja}. After a few settling time scales, an equilibrium will be reached in which the downward flux from sedimentation is balanced by the upward mass flux from diffusion. In this case, the time derivative of the advection-diffusion equation becomes zero and the equation can be integrated analytically. Assuming a vertically isothermal disk and constant diffusivity, \citet{Fromang:2009ja} derived the dust density distribution as
\begin{equation}
\rhod(z) = \rho_{\mathrm{d},0}\,\exp\left[-\frac{\St_\mathrm{0}}{\alpha} \left( \exp\left( \frac{z^2}{2\,\Hg^2}\right)-1\right)-\frac{z^2}{2\,\Hg^2}\right],
\label{eq:sett_mix_solution}
\end{equation}
where a subscript of 0 denotes mid-plane values. Close to the mid-plane, this profile indeed approaches a Gaussian profile with the scale height estimated above, as can be seen by letting $z\ll H$ in \cref{eq:sett_mix_solution}.

\subsection{Radial Structure}
\label{sec:dust-disk:radial}

\subsubsection{Particle Sizes}
\label{sec:dust-disk:radial:sizes}
\begin{figure}[htp]
\begin{center}
  \includegraphics[width=\hsize]{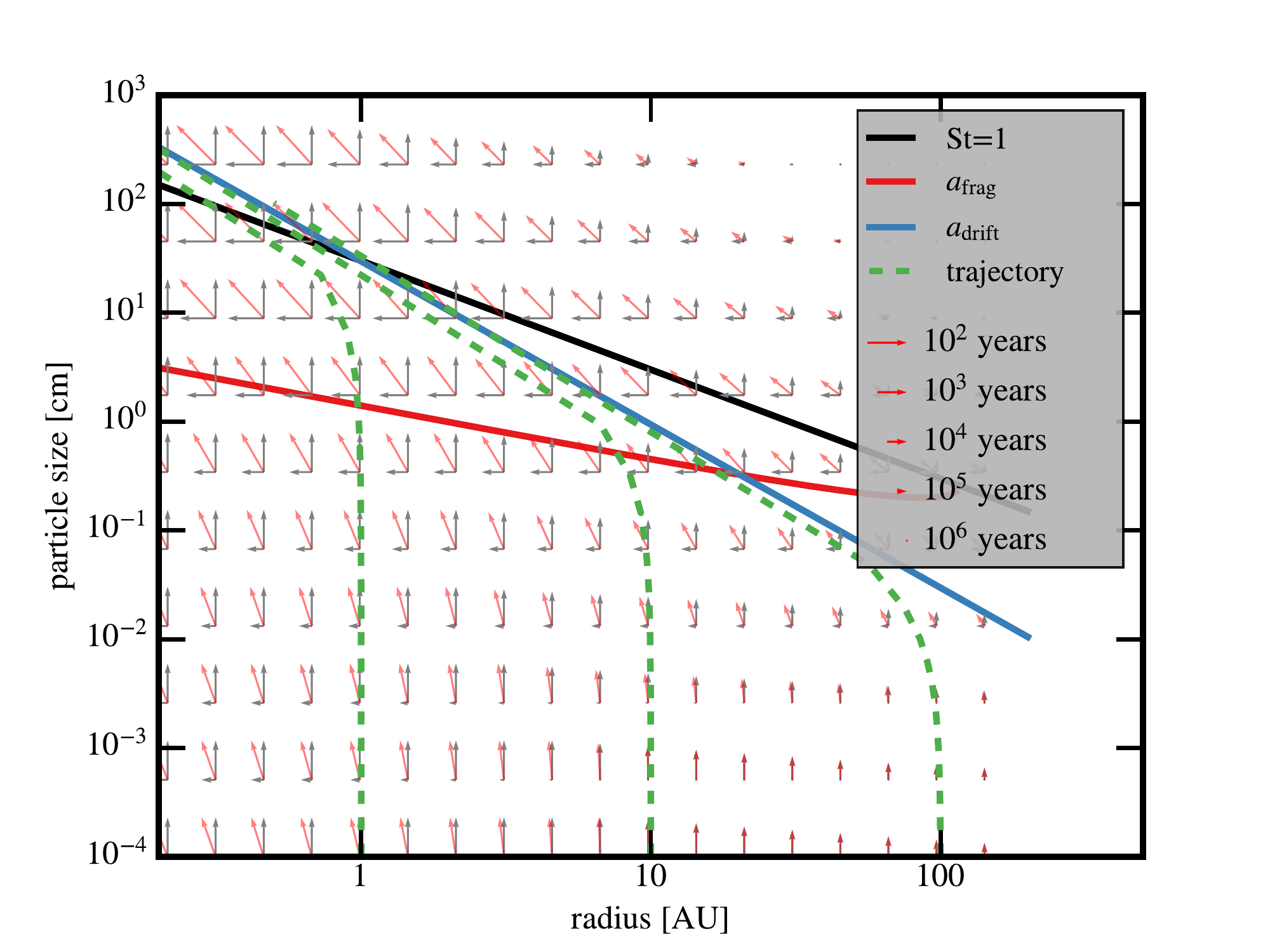}
  \caption{
  Comparison of growth and drift time scales. Disk properties follow the model of \cite{Birnstiel:2014cb} but using a surface density profile of $\Sigg = \SI{0.28}{g.cm^{-2}}\, r_\mathrm{d}/r$ for $r<r_\mathrm{d}$ and $r_\mathrm{d} = \SI{200}{AU}$. The dust-to-gas ratio was chosen to be $\epsilon=5\times 10^{-3}$. The arrow length is inversely proportional to the logarithmic time scales, $r/\dot r$ in the horizontal, and $a/\dot a$ in the vertical direction. Trajectories were calculated for monodispere growth and drift, neglecting fragmentation. Fragmentation limits further growth in the inner $\sim\SI{20}{AU}$. The fragmentation barrier (red line) was calculated for a fragmentation threshold of \SI{10}{m.s^{-1}} appropriate for icy grains. In the inner disk, it will be at smaller sizes due to the reduced fragmentation threshold of silicate grains.
  }
  \label{fig:arrows}
\end{center}
\end{figure}

In the radial dimension, the same mechanisms are at play, dust drift and turbulent mixing. In addition to that, also the radial flow of the gas transports dust particles. A good intuition of the global transport can again be derived by comparing time scales. The growth time scale is
\begin{equation}
t_\mathrm{grow} = \frac{a}{\dot a} = \frac{a\,\rhos}{\rhod\,\Delta v},
\end{equation}
where $a$ is the particle radius and we used \cref{eq:da_dt}. For simplicity, we will assume the relative velocities to follow $\Delta v \simeq \sqrt{3\alpha\St}\cs$ (\cref{eq:v_turbulent} for $\St \ll 1$) and the mid-plane dust and gas densities to be
\begin{equation}
\rho_\mathrm{d/g,0} = \frac{\Sigma_\mathrm{d/g}}{\sqrt{2\pi}\,H_\mathrm{d/g}}.
\label{eq:rho_d_midplane}
\end{equation}
Substituting also the definition of the Stokes number (\cref{eq:stokes-number}) simplifies the growth time scale to
\begin{equation}
t_\mathrm{grow} = \frac{1}{\epsilon\,\Omega},
\label{eq:t_grow}
\end{equation}
where $\epsilon$ denotes the dust-to-gas mass ratio. \cref{fig:arrows} depicts this growth time scale and the time scales for radial motion $r/u_\mathrm{r}$ as arrows, where longer arrows correspond to faster growth/transport.  Let us again imagine all particles to be of the same size (monodisperse growth), initially one micrometer. In the beginning the particles will not drift radially, they will only grow to larger sizes. As the particle size increases, the drift motion will start to become significant and the imagined ``trajectory'' of our dust particle swarm will curve radially inwards. The dashed lines in \cref{fig:arrows} show such ``streamlines'', starting at 1, 10, and 100~AU. Since the growth time scale is proportional to the orbital time scale, we see that growth is slow in the outer regions. Particles in the inner regions will have grown and drifted inwards long before particles in the outer regions have grown significantly. The entire dust disk is therefore sustained by the outermost regions, which act as a reservoir of inflowing mass. The global evolution of the dust disk is thus set by the growth time scales of the outer most disk regions \citep{Garaud:2007ks,Birnstiel:2012ft}. 

\cref{fig:arrows} also shows that all the trajectories converge towards the solid black line, which marks the particle sizes at which the local growth time scale equals the local drift time scale \citep{Klahr:2006iv}, which is called the drift size limit \citep{Birnstiel:2012ft}
\begin{equation}
a_\mathrm{drift} \simeq 0.35 \frac{\Sigd}{\rhos\gamma}\left(\frac{\Hg}{r}\right)^{-2},
\label{eq:a_drift_limit}
\end{equation}
where $\gamma= \left|d\ln P/d\ln r\right|$.

Particles below this curve will grow faster than they drift -- particles below it will drift faster than they grow. The dashed black line denotes the particle size that corresponds to a Stokes number of unity, i.e. the size where drift rates and turbulent collision velocities reach a maximum. In this particular plot, a somewhat lower than canonical dust-to-gas ratio of $\epsilon=5 \times 10^{-3}$ was assumed as evolved disks will have lost dust mass due to drift, or the formation of larger bodies. This also emphasizes the effect that particles not necessarily grow towards $\St=1$. If they did (as would be the case for a larger dust-to-gas ratios), then radial drift would be very fast, and this would very quickly reduce the dust-to-gas ratio.

So far, we neglected fragmentation and bouncing, but as particles grow, their relative velocities increase. Let us suppose particles fragment at impact velocities above a threshold velocity $v_\mathrm{frag}$. Given that turbulent relative velocities increase with size according to \cref{eq:v_turbulent}, we can derive a maximum size particles can reach before impact velocities become larger than the fragmentation threshold. We find that this size is
\begin{equation}
a_\mathrm{frag} \simeq 0.08\,\frac{\Sigg}{\rhos\,\alpha}\left(\frac{v_\mathrm{frag}}{\cs}\right)^2.
\label{eq:a_frag_limit}
\end{equation}  
Only if the highest turbulent mean velocity $v_\mathrm{rms} = \sqrt{3\alpha/2}\cs$ are smaller than $v_\mathrm{f}$, fragmentation is not happening and \cref{eq:a_frag_limit} does not apply \citep[but see][for the effect of taking the distribution of velocities into account]{Windmark:2012bg}. \citet{Birnstiel:2012ft} found that fragmentation tends to dominate in the inner regions of the disk, while outer regions, or regions of lowered dust-to-gas ratio are limited by radial drift. \Cref{fig:arrows} shows that inside of about 20~AU, fragmentation (the red line) will limit particle growth, while outside of this radius, the drift limit applies. The trajectories in \cref{fig:arrows} ignore the effects of fragmentation -- particles are limited to the fragmentation barrier (\cref{eq:a_frag_limit}) there, and will therefore drift at a lower rate as compared to particles at the drift limit.

This picture of growth, drift, and fragmentation may be quite simplistic, but it was found to match well the behavior seen in numerical simulations \citep[see][]{Birnstiel:2012ft}, which incorporate many more details. A similar behavior, at least in the outer disk was also found in cases where the porosity of the grains was allowed to change: \citet{Okuzumi:2012kd} considered icy grains and assumed that they do not fragment\footnote{Water-ice particles fragment only at velocities beyond $v_\mathrm{f}\gtrsim 10$~m~s$^{-1}$, see \citet{Gundlach:2015iu}; silicate or carbonaceous grains fragment at velocities $\lesssim 1$~m~s$^{-1}$, see \citet{Blum:2008fi}.}. In their simulations, particles grow in a fractal way reaching very low filling factors ($\sim 10^{-5}$). Such particles in the inner disk could grow much quicker and thus possibly overcome the drift size limit (\cref{sec:planetesimal-formation:collisional:porosity}). According to \citet{Krijt:2015bu}, however, erosion is expected to limit growth beyond Stokes numbers of unity. But all these simulations find that particle growth beyond about 10--20~AU is limited by the radial drift motion of the grains, irrespective of their porosity.

\subsubsection{Dust Surface Density Profiles}
\label{sec:dust-disk:radial:densities}

Once we have derived the maximum particle size as function of distance to the star $a_\mathrm{max}(r)$ (it could be either due to fragmentation, drift, or other size-limiting effects), we can use this to derive an expected dust surface density profile as follows. Practically all particle size distributions have most of the mass in the largest grains (interstellar medium (ISM): \citealp{Mathis:1977hp}, circumstellar disks: e.g., \citealp{Brauer:2008bd,Birnstiel:2012ft,Zsom:2011fq,Okuzumi:2012kd}, debris disks: e.g., \citealp{Dohnanyi:1969dw}). In protoplanetary disks, also the transport velocity of grains is largest for the largest grains (as long as $\St\lesssim 1$). This means that the total mass flux (density times velocity), the rate at which the dust is transported is dominated by the largest grains. We can now simply assume that all the mass is near the maximum particle size $a_\mathrm{max}(r)$, and calculate the dust transport velocity of those particles $v_\mathrm{r}$. Furthermore, we saw that the mass flux is set by the outer disk regions, which slowly ``leak'' dust inwards and this dust mass flow is conserved, so the mass accretion rate should approach a constant value. This allows us to write the dust mass accretion rate as
\begin{equation}
\dot M_\mathrm{d} = 2\,\pi\,r\,\Sigd\,v(r) = const.,
\end{equation} 
and solve for the dust surface density profile
\begin{equation}
\Sigd(r) = \frac{\dot M_\mathrm{d}}{2\,\pi\,r\,v(r)}.
\label{eq:sig_d_steady}
\end{equation}
Hence, different prescriptions of $a_\mathrm{max}(r)$ will yield different dust surface density profiles \citep{Birnstiel:2012ft}.
For typical assumptions ($\Sigg\propto r^{-1}$, $T\propto r^{-0.5}$), \cref{eq:sig_d_steady} predicts $\Sigd\propto r^{-0.75}$ in the drift limit (applicable in the outer disks) and a steeper slope of $\Sigd\propto r^{-1.5}$ in the inner parts of the disk, where fragmentation dominates \citep{Birnstiel:2012ft}. This is in agreement with observations of \citet{Andrews:2012ii} and \citet{Menu:2014hg}.   
Observational constraints on $a_\mathrm{max}(r)$ and $\Sigd(r)$ thus enable us to probe the collisional processes in protoplanetary disks (see \cref{sec:observations:results}).

The fact that the growth time scales become longer in the outer regions means that the dust surface density is draining from the inside-out, where the mass flux is supplied by regions lying further and further outside. In this framework \citep[see,][]{Garaud:2007ks,Birnstiel:2012ft}, particle growth can be interpreted as a radially expanding ``pebble formation'' front, which allows the dust mass accretion rate to be estimated analytically \citep{Lambrechts:2014iq}. As \citet{Ormel:2010ii} have demonstrated, pebbles can efficiently be accreted by growing proto-planets as the drag force around them effectively increases their cross section. With growth and drift providing a constant supply of inward drifting grains of the right sizes to be accreted, the growth time scale of cores, particularly in the outer regions of disks can be reduced significantly \citet{Lambrechts:2012gr,Lambrechts:2014iq}.

At this point, it should be mentioned that the overall grain sizes and surface density profiles seem to be in reasonable agreement with observations (see \cref{sec:observations:results:density-profiles-disk-and-hole-sizes,sec:observations:results:spectral-indices-grain-sizes}).  However the radial drift mechanism seems too efficient -- for typical disk parameters, and particles of millimeter sizes, the expected disk life times are at least one order of magnitude too short \citep{Brauer:2007dd}, as compared to observational constraints (see \cref{sec:observations:results:disk-lifetimes}).

\section{Planetesimal Formation}
\label{sec:planetesimal-formation}
The previous sections have already achieved the first step towards planet formation: the initial growth from sub-micrometer sized dust grains towards macroscopic solid particles. However this is only the first step of many. The next step is the formation of the building blocks of planets, so called planetesimals. They are defined as being bound by their own gravitational attraction (as opposed to their surface and material binding forces), which typically happens at sizes above several kilometers \citep[e.g.,][]{Benz:1999cj}. How planetesimal are assembled into planets is the subject of the next chapter in this book. In the following sections, we will discuss the step from dust to planetesimals, i.e., from particles that are strongly affected by gas drag to large bodies that are basically unaffected by gas drag. We will start by reviewing collisional planetesimal formation, i.e. particles growing successively larger via collisions  in \cref{sec:planetesimal-formation:collisional}. \cref{sec:planetesimal-formation:gravitational} will then discuss various ideas how planetesimals form via gravitational instabilities and \cref{sec:planetesimal-formation:synthesis-models} will review recent synthesis models. 

\subsection{Collisional growth}
\label{sec:planetesimal-formation:collisional}
The basic physics of collisional growth was already briefly discussed in \cref{sec:dust-growth}: particle collisions are driven by relative velocities between grains (\cref{sec:dust-growth:impact-velocities}). The collision rates depend on the relative velocity, but also the number of colliding particles as well as their cross sections. Assuming perfect sticking upon impact and compact particles, we defined a growth time scale (\cref{eq:da_dt}) and already saw that the fast inward drift tends to be more effective in removing particles than particle growth can grow them. At the same time, collision velocities of larger grains can become too large and fragmenting collisions potentially prevent further growth. In the following, we will discuss how these obstacles could be overcome.

\subsubsection{Compact growth}
\label{sec:planetesimal-formation:collisional:compact-growth}
\cref{fig:windmark} shows the collisional outcome for conditions at 1~AU (where water is not frozen out on the grain surfaces) according to the model of \citet{Windmark:2012gi}. For collisions of equal-sized particles (the diagonal of \cref{fig:windmark}), we see that as particles grow, their sticking probability decreases and bouncing becomes the dominant collisional outcome. The situation where particles stop growing and become stuck at the sticking/bouncing transition was termed the bouncing barrier \citep{Zsom:2010hg}. In \cref{fig:windmark}, this is approximately below $10^{-2}$~cm.

\cref{fig:windmark} also shows that even if the bouncing barrier could be overcome, collisions of equal-sized particles will become more violent: fragmentation and erosion come into play \citep[e.g.][]{Weidenschilling:1997im,Dullemond:2005hf,Brauer:2008bd}, limiting further growth at the fragmentation barrier \citep[e.g.][]{Birnstiel:2009hc}. However, there exists a pathway towards growing larger bodies: suppose even only a single particle exists that has a size of about a centimeter, then its collisions with the small particles at the bouncing barrier lie in the green region of \cref{fig:windmark}, where ``mass transfer'' leads to a net growth of the larger particle. The effect is similar to throwing a snow ball against a wall: the impactor is fragmented but still able to deposit some of its mass onto the target. This type of collisional outcome was first seen in laboratory experiments by \citet{Wurm:2005kv} and subsequently in \citet{Paraskov:2007hm}, \cite{Teiser:2009gn} as well as in \citet{Kothe:2010bd}. \citet{Windmark:2012gi} showed that growth of larger bodies is possible in such a way, but it is slow: 10~m sized boulders could grow within about $2\times 10^{5}$~years, 100~m sized boulders in about $10^6$~years.

The question where these ``seed particles'' beyond the bouncing barrier could come from was answered in \citet{Windmark:2012bg} \citep[see also][]{Garaud:2013ja}: the results so far assumed that two particles of a given size always collide with their \textit{mean} relative velocity. However, particle velocities and thus also their relative velocities will have a non-zero dispersion around the mean. Thus, two particles whose mean collision outcome might be bouncing or fragmentation have in fact a non-zero probability of colliding either at high velocity (causing fragmentation) or a low velocity (causing sticking). This way, some particles might be ``lucky'' enough to avoid disruptive collisions.

But this is both a cure and a curse. Let us consider the positive effect first: ``lucky'' particles can continue to grow into regions where the mean collision outcome would not allow growth. The bouncing barrier thus becomes permeable: particles either bounce (= no growth, but no destruction) or they grow\footnote{The onset of fragmentation is still so  far away in velocity space that the probability of a fragmenting impact velocity is negligibly low.}. Thus, the bouncing barrier will only slow down growth, but it will not stop it. Particles will continue to grow beyond the bouncing barrier, as found by \citet{Windmark:2012bg}.

The negative effect is that also fragmentation becomes relevant already at smaller sizes than before. This also slows down growth and unlike bouncing, fragmentation leads to a reduction of the particle size. For a particle to become twice as massive as the fragmentation limit $a_\mathrm{frag}$, one ``lucky'' collision with an equal sized particle is enough. To double its size, it already needs 8 lucky collisions, and so on. The chances of the particle to grow become smaller and smaller. Fortunately, there are many particles: for the minimum mass solar nebula \citep{Weidenschilling:1977kq} at 1~AU, a region of width $\Hg$ would contain a mass corresponding to $10^{29}$~particles of millimeter sizes. We could therefore expect some particles to be ``lucky enough'' to continue to grow to larger sizes. But still: only a very small fraction of the entire mass would be in those large bodies -- if there were more of them, collisions among them would be frequent and destroy them. And finally, as discussed above the growth time scale for sweep-up growth are long, longer than the radial drift time scale, so particles would still drift inwards before they could grow to larger sizes.

\subsubsection{Porosity evolution}
\label{sec:planetesimal-formation:collisional:porosity}
A solution to this might lie in the internal structure of the particles: so far, we considered particle growth inside the snow line, where the constituents of dust aggregates, the so called \textit{monomers} consist of silicates or carbonaceous particles. These particles are sticking to each other due to the weak van der Waals force. Further outside, where temperatures in the disk are lower, water will freeze out on the grain surfaces. Water ice develops hydrogen bounds leading to surface energies that are about 10 times higher \citep{Heim:1999hn,Gundlach:2011jj}. These stiffer bounds between monomers might help overcome the radial drift barrier or destructive collisions: \citet{Wada:2008eh} simulated collisions of icy aggregates of sizes of a few micrometers\footnote{These results strongly depend on the monomer size. These works assumed a monomer size of \SI{0.1}{\um}.} and derived a fragmentation threshold velocity of around \SI{50}{m.s^{-1}}. This is indeed higher than the $\sim$ \si{m.s^{-1}} measured for silicate dust \citep[e.g.,][]{Wurm:1998kg}, as was expected, but recent laboratory studies found threshold velocities that were not quite as high, with about \SI{10}{m.s^{-1}} \citep{Gundlach:2015iu}. Still, higher threshold velocities of icy aggregates allow growth to proceed to larger sizes or possibly avoid destructive fragmentation completely.

As another consequence of strong surface energy, compression of aggregates becomes more difficult. This prevents bouncing and the bouncing barrier \citep{Wada:2011gd,Seizinger:2013br} and means that  aggregates become more porous. For particles in the Epstein drag regime, the growth time scale does not depend on the surface-to-mass ratio (see \cref{eq:t_grow}). But in the inner regions of disks, where gas densities are high, particles may be in the Stokes drag regime. In this case, the growth time scale at a given Stokes number becomes inversely proportional to the particle size \citep{Okuzumi:2012kd}. At the crucial regime around Stokes numbers of unity, extremely fluffy particles (internal densities of $\ll~\SI{1e-3}{g.cm^{-3}}$) have a very large size resulting in very short growth time scales, short enough to overcome the radial drift barrier \citep{Okuzumi:2012kd,Kataoka:2013kx}. This mechanism could thus form relatively compact planetesimals by avoiding the growth barriers via extremely low internal densities, followed by compression via gas drag and self-gravity to reach cometary densities \citep{Kataoka:2013kx}.

There are however some caveats: first, the gas densities have to be high enough for particles to be in the Stokes drag regime, which is only expected to be the case inward of $\sim$10~AU in the model of \citet{Okuzumi:2012kd}. Second, sintering can cause the break-up of the aggregates. The sintering time scales are shorter than the growth time scales inside of a few~AU \citep{Sirono:2011be,Sirono:2011cv}, preventing the formation of planetesimals in such a model. Finally, \citet{Krijt:2015bu} have considered the additional effect of erosion, which was not included in the study by \citet{Okuzumi:2012kd} and \citet{Kataoka:2013kx}. The relative velocity between unequal sized particles can be much larger than between equal sized particles (see~\cref{fig:rel_vel}). Even if equal size particles do not destroy each other, small and large bodies can collide at velocities above the erosion threshold velocity. If this is the case, erosion stops further growth at Stokes numbers around unity also resulting in somewhat more compact aggregates \citep{Krijt:2015bu}. Recently measured erosion threshold velocities of around 15~m~s$^{-1}$ \citet{Gundlach:2015iu} seem to support this result.

Even if erosion prevents the collisional growth of planetesimals, not all is lost: particles are still expected to grow to Stokes numbers large enough to participate in gas induced particle concentrations which can lead to gravitationally bound bodies, as we will discuss in the following.

\subsection{Formation of planetesimals by gravitational collapse}
\label{sec:planetesimal-formation:gravitational}

As discussed in \cref{sec:dust-dynamics}, growth of dust aggregates also changes the aerodynamic properties of the aggregates. The increased stopping time leads to a partial decoupling of the motion of the dust aggregates from the motion of the gas and this in turn allows particles to sediment to the mid-plane of the protoplanetary disc, drift radially towards the central star and to become concentrated in the turbulent gas flow. In the following, we will discuss some of the dynamical effects that are able to cause strong local concentrations of dust particles. We will not discuss here the turbulent concentration mechanism proposed by \citet{Cuzzi:2008js}, where chondrule-sized particles are concentrated on the smallest scales of the turbulent flow, as the relevance of this mechanism for protoplanetary discs is still under exploration. We refer the reader to the recent review paper by \citet{Johansen:2015tc}, and references therein, for a detailed description of the latest development in the study of small-scale turbulent concentration.

\subsubsection{Pressure bumps}
\label{sec:planetesimal-formation:gravitational:pressure-bumps}

One of the most generic flow structures that can concentrate particles is the so-called pressure bump. Pressure bumps are axisymmetric overpressure regions that prevail in perfect balance between the outwards-directed pressure gradient force and the inwards-directed Coriolis force. Particles do not react to the pressure gradient directly, but the surrounding zonal flow envelope is super-Keplerian on the inside of the pressure bump and sub-Keplerian on the outside. Dust particles sense a tailwind on the inside of the bump and a headwind on the outside of the bump, resulting in their migration to the centre of the bump (cf. \cref{eq:radial_drift}). The convergence region will move slightly inwards towards the star in the presence of a global, radial pressure gradient.

Pressure bumps can arise spontaneously in protoplanetary discs by differential transport of angular momentum. While simplified models of protoplanetary accretion discs with a constant value of the turbulent viscosity coefficient $\alpha$ always display outwards transport of angular momentum, any variation in the strength of the turbulence can lead to a local pile-up of angular momentum -- this is the zone that in turn forms the pressure bump through the convergence of the radial flow induced by the Coriolis force.

Simulations of turbulence driven by the magnetorotational instability \citep{Balbus:1991fi} show that large-scale variations in the strength of the turbulence lead to the spontaneous formation of pressure bumps at the largest scales of the turbulent flow \citep{Johansen:2007cl,Simon:2012dq,Dittrich:2013ix,Simon:2014dr}. These pressure bumps are quasi-stable and live for hundreds of orbits before they disassemble and reform. Particle concentrations inside pressure bumps reach values above the Roche density, and very large planetesimals can form by the subsequent gravitational contraction and collapse phases \citep{Johansen:2011jo}.

Turbulence is nevertheless a two-edged sword. Collision speeds between m-sized particles moving in turbulence caused by the magnetorotational instability reach tens to hundreds of meters per second. Hence it is not clear that fully developed turbulence is a good environment for forming planetesimals -- despite the emergence of pressure bumps that concentrate particles.

The differential angular momentum transport needed to form pressure bumps can also arise from radial variations in the very nature of the turbulence. The transition from regions of fully developed turbulence caused by the magnetorotational instability and the ``dead zone'' whose ionization fraction is too low to sustain the magnetorotational instability has been shown to lead to the formation of pressure bumps at the inside and outside of the dead zone \citep{Lyra:2008gu,Lyra:2009ga,Kretke:2009ex}. Such sweet spots could experience planetesimal formation ahead of the bulk parts of the protoplanetary disc and hence have important implications for the formation sequence of planets in planetary systems.

\subsubsection{Vortex trapping}
\label{sec:planetesimal-formation:gravitational:vortex-trapping}

Azimuthally elongated vortices may form directly from a flow instability, such as the baroclinic instability, or result from Rossby wave instabilities that are triggered by high-amplitude pressure bumps.

The baroclinic instability is a process analogous to radial convection \citep{Klahr:2003kq,Lesur:2010jn,Raettig:2013cn}. The non-linear state of the instability is characterized by the emergence of large-scale, slowly overturning vortices. Such vortices can trap dust particles \citep[e.g.,][]{Barge:1995vd,Meheut:2012di,Lyra:2013ix,Raettig:2015vc} in a similar way to pressure bumps \citep{Paardekooper:2004bf,Pinilla:2012ke,Zhu:2012kr,Birnstiel:2013es,Lyra:2013ix}. Recent observations show a remarkable resemblance with expected signatures of these models (see \cref{sec:observations}). Dust feedback (the impact of the dust dynamics on the gas dynamics) nevertheless limits the particle concentration obtainable inside such vortices, as the vortex flow is destroyed by particle feedback for a dust-to-gas ratio above unity \citep{Johansen:2004jp,Fu:2014gk}. Still, the achieved over densities in vortices might trigger planetesimal formation via the streaming instability \citep{Raettig:2015vc}, which will be discussed in the following section. 

\subsubsection{Streaming instability}
\label{sec:planetesimal-formation:gravitational:streaming-instability}

The sedimentation of dust particles paves the way for a concentration mechanism in which the particles play an active role through their friction on the gas. As the dust-to-gas ratio in the mid-plane approaches unity, the back-reaction friction force from the particles onto the gas becomes strong enough to accelerate the gas to orbit at closer to the Keplerian speed together with the particles. Thus the headwind on the particles is reduced and the particles fall more slowly towards the star. An overdense structure in the mid-plane (such as an axisymmetric particle filament) would fall even more slowly than the surrounding particles. This is now a run-away situation because these faster-drifting particles continue to drift at their full rate and thus pile up in the filament -- this is the essence of the so-called streaming instability \citep{Youdin:2005de}.

The characteristic length scale of the streaming instability is approximately 5\% of a gas scale-height $H$, although the separation between filaments in the non-linear state is higher, close to 0.2 gas scale-heights \citep{Yang:2014ix}.

Particle concentration by the streaming instability can reach at least several thousand times the local gas density \citep{Bai:2010gm,Johansen:2012kr}. The particle filaments appears to obtain a fractal structure where smaller-scale filaments embedded within the main filament allow very high particle densities at the smallest, resolvable scales of the simulations \citep{Johansen:2012kr}. The high mass loading in the gas could be an additional factor needed to reach very high densities, as the turbulent diffusion of the gas is drastically reduced in such particle dominated flows \citep{Johansen:2009jf}.

\begin{figure}[htp]
\begin{center}
  \includegraphics[width=0.8\hsize]{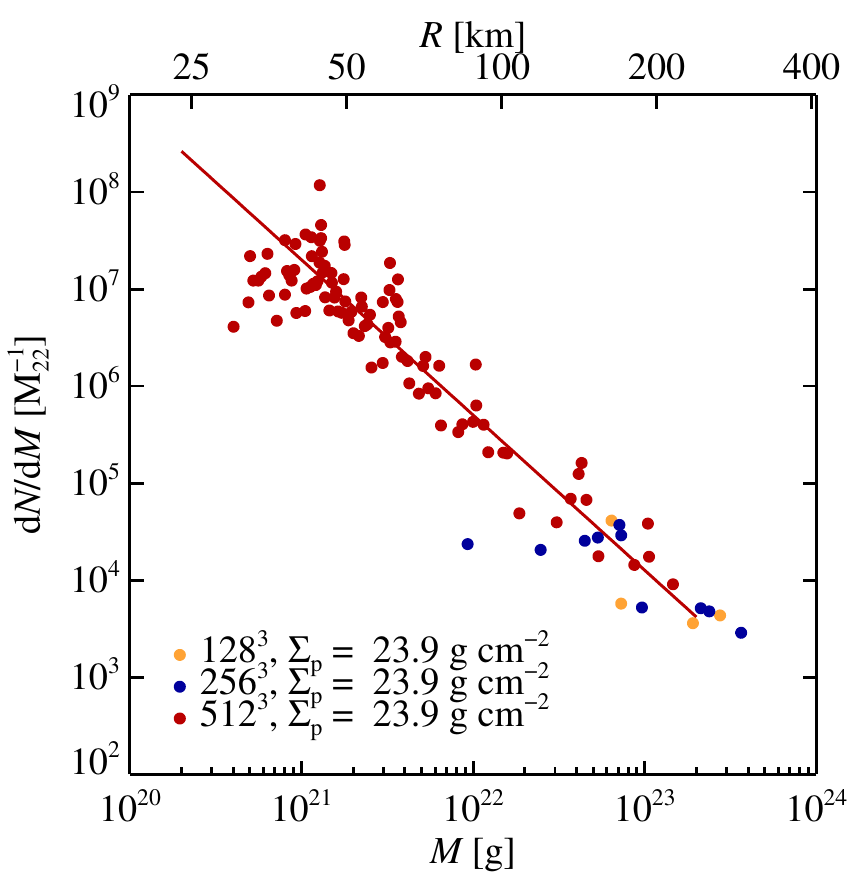}
  \caption{Differential mass distribution of planetesimals formed by the streaming instability at three different numerical resolutions, from \citet{Johansen:2015he}, as a function of planetesimal mass (lower axis) and size (upper axis). The mass distribution is given per \SI{1e22}{g} in an annulus of \SI{1}{AU} width situated at \SI{2.5}{AU} from the central star. A fit of ${\rm d}N/{\rm d} M \propto M^{-1.6}$ is overplotted (full line). Such a differential mass distribution is dominated in mass by the few largest bodies, but the number is dominated by the smallest bodies. The results here are for a column density of approximately six times the value in the minimum mass solar nebula; reducing the column density to the nominal value leads to a characteristic planetesimal size closer to 100 km in radius \citep{Johansen:2015he}.}
  \label{fig:size_distribution}
\end{center}
\end{figure}

The characteristic planetesimal size forming by the gravitational collapse of the over-dense regions formed by the streaming instability depends sensitively on the column density of the particles. Models with a few times the minimum mass solar nebula worth of particles in the asteroid belt give rise to Ceres-mass planetesimals \citep{Johansen:2012kr}, while the characteristic planetesimal size falls with decreasing particle column density, down to 100 km for the more realistic case where the column density is slightly lower than in the minimum mass solar nebula (see \cref{fig:size_distribution}).

An important prerequisite for the formation of dense particle filaments by the streaming instability appears to be a sufficiently high ratio of particle column density to gas column density, $Z=\Sigma_{\rm p}/\Sigma_{\rm g}$. Strong particle concentration is triggered above a threshold value of $Z\approx0.015$ \citep{Bai:2010gm}. This threshold is approximately constant for particle sizes from St=0.03 to St=0.3 (corresponding to approximately cm-dm sized particles in the asteroid belt and to mm-cm sized particles in the outer Solar System). Smaller and larger particles require a higher mass-loading to trigger particle concentration \citep{Youdin:2005de,Carrera:2015hz}.

\subsection{Synthesis: combined models}
\label{sec:planetesimal-formation:synthesis-models}

Clearly dust coagulation and particle concentration are in no way mutually exclusive processes. In fact, the two are intimately coupled. Dust growth to mm and cm sizes is key for allowing particles to decouple from the gas and be concentrated, while the dense environment caused by particle concentration incites high collision rates, where the outcomes of the collisions in turn determine the continued contraction of the cloud.

The growth of dust aggregates to sizes that can undergo concentration by the streaming instability is a subject of intense scrutiny at the moment. \citet{Drazkowska:2014dj} showed that the appropriate cm-sized particles can form (at least outside of the ice line where sticking is easier), but that production of sufficient amounts of large dust aggregates to trigger the streaming instability requires a super-solar metallicity (or photoevaporation of the gas). The challenge is that only a fraction of the total dust mass reaches the particle sizes necessary to trigger the streaming instability and that the streaming instability becomes self-limiting by removing the dust mass and lowering the metallicity below the necessary values. Similar conclusions were reached by \citet{Krijt:2016jm} for the growth of very fluffy ice particles. It thus appears that there remains a gap in our understanding between the amount of large particles that can be produced by direct growth and the amount needed to trigger strong particle concentration.

Pebbles in the asteroid belt appear to be represented by the mm-sized chondrules found in primitive meteorites. Typical chondrules have Stokes numbers around 0.001 when assuming a background gas density like the minimum mass solar nebula. Sedimentation of such small particles to the mid-plane is not very efficient, as even a moderate degree of turbulent stirring is enough to lift them up to the disk atmosphere. Particle concentration by the streaming instability also appears problematic for mm-sized particles in the asteroid belt \citep{Carrera:2015hz}. The situation would be better if the gas density was lower than in the minimum mass solar nebula, as motivated by observations \citep[e.g.][]{Andrews:2012ii}. A reduction in gas density by a factor of 10 would increase the friction time correspondingly and put chondrules within the range of particles that can be concentrated by the streaming instability. It is also possible that chondrules formed chondrule aggregates, the sticking facilitated by acquiring porous dust rims \citep{Ormel:2008bn}. Alternatively, other hydrodynamical concentration mechanisms may be are able to bridge the gap, e.g. the baroclinic instability/convective overstability \citep{Klahr:2003kq} or the vertical shear instability \citep{Nelson:2013cd}. Such instabilities do not necessarily rely on an initially high dust-to-gas ratio to further concentrate the dust. The concentrations formed by these mechanisms could then trigger the streaming instability even in cases where the overall metallicity would be too low. 

Hydrodynamical simulations of planetesimal formation by gravitational collapse typically employ a fixed grid for the dynamics and hence are not able to follow the collapse of the particle clouds down to planetesimal sizes. This collapse phase is nevertheless so decoupled from the remaining disc that it can be modelled in a separate N-body simulation. \citet{Nesvorny:2010da} showed that the collapse typically leads to the formation of binary planetesimals, in good agreement with the prevalence of binaries in the classical cold component of the Kuiper belt \citep[e.g.,][]{Noll:2008bj}. \citet{WahlbergJansson:2014ku} included a detailed particle collision model in the collapse and found that a large fraction of the pebbles survive the collapse without undergoing fragmentation, leading to the formation of pebble-pile planetesimals consisting of primordial pebbles from the protoplanetary disc.

This ``pebble paradigm'' for planetesimal formation will be put further to the test in the future with the observations of minor bodies in the Solar System. The presence of ``goosebumps'' on the comet 67P has been interpreted as the primordial pebbles of the solar protoplanetary disc \citep{Sierks:2015db} -- although if this interpretation is true, then that poses a challenge for coagulation theory and experiments to explain how the dominant pebble sizes can be more than one meter, in contrast to drift dominated models that predict sizes closer to 1 cm in the outer disc \citep{Birnstiel:2011ks}.

\section{Observations}\label{sec:observations}

\subsection{Methods}\label{sec:observations:methods}
The dust particles in disks span a large range of temperatures from over one thousand to only tens of K, decreasing with the distance to the central star. The grains emit thermal emission from the near-infrared band where the hot inner disk dominates, through the mid-infrared where the warm disk atmosphere contributes most, to far-infrared and \submm wavelengths where the optical depth is lower and we can see deeper into the disk. Apart from the inner $\sim$\SI{10}{AU}, disks are normally optically thin in the millimeter continuum and the emission thus probes the cool disk interior where most of the mass resides.

Various techniques have been developed to study disks based on dust emission. The infrared imaging of star-forming regions of young clusters is one of the most efficient ways to characterize disks properties of a large sample of young stars with disks. Our understanding of disk structure and evolution has been advanced progressively with the advent of various space infrared telescopes, e.g. IRAS, ISO, Spitzer, and Herschel, and complemented with a lot of ground-based telescopes with infrared instruments. From these observations, in combination with \submm imaging, the
entire spectral energy distributions (SED) of disks can be constructed. Modeling the SEDs of disks has become a classical way to investigate general disk properties. However, because of the lack of spatial resolution, the results from SED modeling are strongly model dependent. The resolved images from \submm interferometric observations in dust continuum emission and molecular line emission can constrain the surface density distribution of dust and gas, as well as dust grain sizes, disk inclinations,  et cetera. A combination of modeling both SEDs and resolved images from \submm interferometric observations has proven to be the best way to study disks \citep{Andrews:2009jo}.

\subsubsection{Infrared Observations}\label{sec:observations:infrared-observations}

\begin{figure}[htp]
\begin{center}
  \includegraphics[width=\hsize]{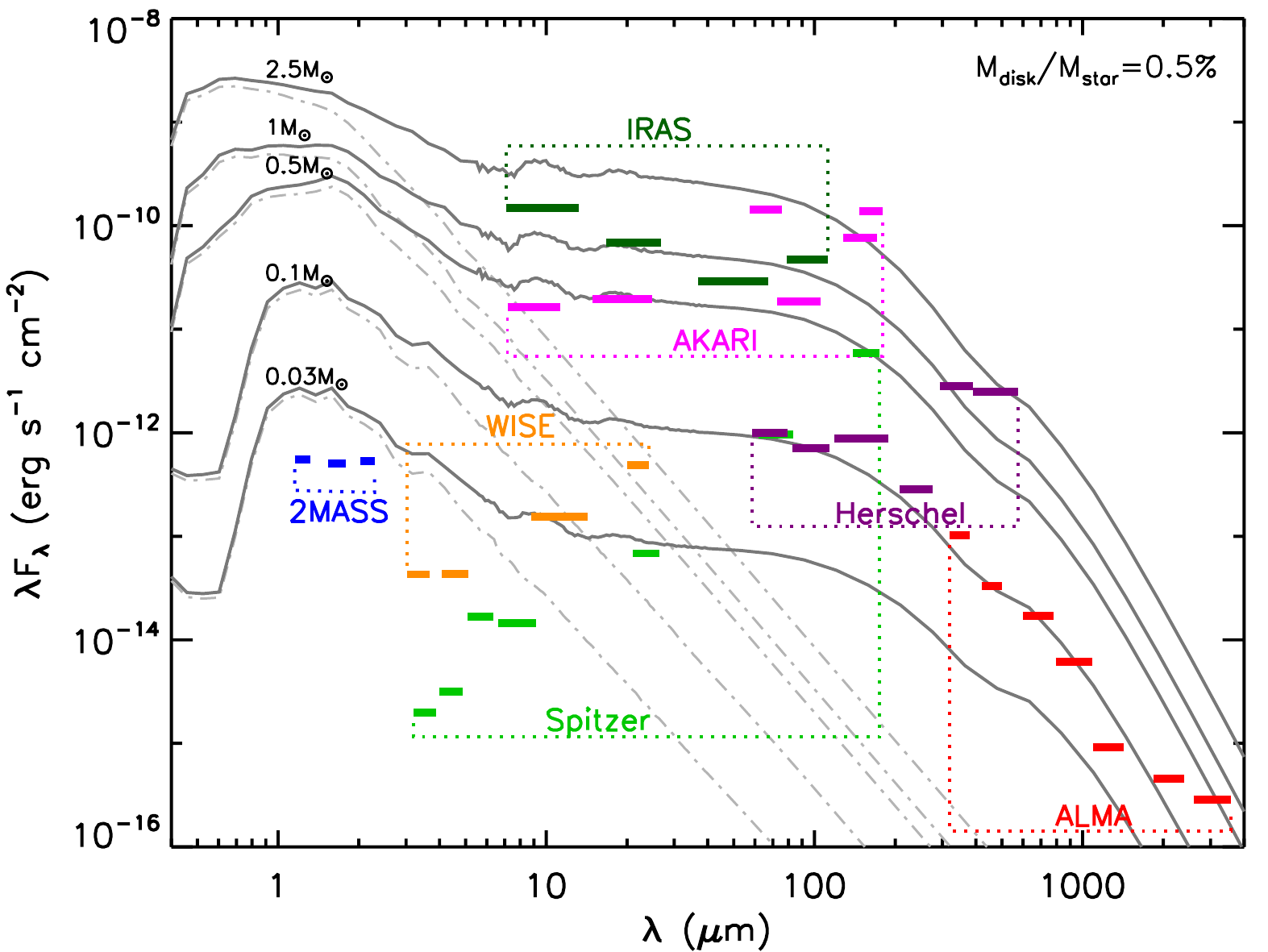}
  \caption{The 3$\sigma$ detection limits of various instruments overlaid on the model SEDs of young disk systems, located at a distance of 300\,pc, with stellar masses of 2.5, 1.0, 0.5, 0.1, and 0.03\,$M_{\odot}$, and a disk-to-star mass ratio of 0.5\%. The SEDs are calculated with  a Monte Carlo radiative transfer code RADMC-3D \citep{Dullemond:2012vq}. The detection limits of IRAS, 2MASS, AKARI, and WISE are typical values for these all-sky surveys. The ones of Spitzer, Herschel, and ALMA are for observations with on source integration time of 1\,minute.}
  \label{fig:SED-sensitivity}
\end{center}
\end{figure}

\citet{MendozaV:1966de,MendozaV:1968gp} presented the first discovery of near- and mid-infrared excess emission of T~Tauri stars. At that moment it was unclear where the excess emission came from. The launching of IRAS in 1983 extended infrared observations to the far-infrared bands. Observations with IRAS revealed that many T Tauri stars, showing near- and mid-infrared excess, also show significant far-infrared excesses, indicating the existence of cold circumstellar dust \citep{1985AJ.....90.2321R}. Various models have been proposed to explain the infrared excess
emission of T~Tauri stars. An optically thick but physically thin circumstellar disk was believed to be the best model for explaining these observations \citep{1987ApJ...319..340M,Adams:1987gy}. ISO, launched in 1995, opened a new infrared window for spectroscopy and discovered the presence of crystalline silicates in disks \citep{1996A&A...315L.245W}. Thanks to the incredible sensitivity of the Spitzer Space Telescope, our knowledge of disks has been significantly improved in the last ten years. Due to the high efficiency in observations, many star-forming regions have been surveyed with Spitzer in both imaging and spectroscopic modes, which advanced our understanding of disk lifetimes, disk evolution with local environments, dust mineralogy in disks around different types of young stars from intermediate mass stars to brown dwarfs \citep{2005Sci...310..834A,Furlan:2006bb,SiciliaAguilar:2006kj,2007ApJ...660.1532B,Hernandez:2007dy,2009A&A...496..453G,SiciliaAguilar:2011cp,Kraus:2012dh,2012A&A...539A.119F,2013A&A...549A..15F}. The Spitzer imaging survey also largely extended the sample of "transition disks". This class of disks (first seen by \citealp{Strom:1989ig}) have optically thin, dust depleted (or even dust-free) inner regions. They were thought to be in the act of clearing the disk from the inside out, therefore called transition disks \citep{Marsh:1993ip,Fang:2009kj,2010ApJ...712..925C,Muzerolle:2010es,2012ApJ...750..157C,Fang:2013br,Espaillat:2014hh}. Recently, the Herschel space telescope, launched in 2009, has presented incredible sensitivity at far-infrared wavelengths, which makes it possible to study dust contents of large samples of disks and to investigate the disks around brown dwarfs at these wavelengths \citep[][and also see sensitivities of various instruments in Fig.~\ref{fig:SED-sensitivity}]{2012A&A...544A..78M,2013PASP..125..477D,2012ApJ...755...67H}.

\subsubsection{SED Modeling}\label{sec:observations:sed-modeling}

Before circumstellar disks could be spatially resolved, the observed infrared excess of young stars was explained using simple disk models. Modeling the observed SEDs became a classical way to characterize global properties of disks. Initially simple geometrically thin and optically thick disks were used in SED modeling \citep{Adams:1987gy,1987ApJ...323..714K}, which was refined by including  an optically thin upper layer above the simple disk models \citep{Chiang:1997ev}. The SED modeling was further improved by 2D/3D radiative transfer disk models  in which the temperature and vertical structure were calculated self-consistently \citep{2003ApJ...598.1079W,2013ApJS..207...30W,Dullemond:2004iy,Dullemond:2012vq}. Even web-based SED fitting tools, built upon those codes are available \citep{Robitaille:2006cb,2007ApJS..169..328R}. In the typical SED modeling approach, the dust properties and disk geometry are parameterized. The SEDs computed from these models are then compared against the observed SEDs to find the best-fitting parameters. The advantage of SED modeling is that one can investigate a large sample of disks since the SEDs can be easily constructed from various imaging surveys. The physical properties of disks can be inferred using this technique and studied statistically, rather than simply quantifying the SED shapes. However, the disadvantage of this technique is that the results are strongly model dependent and highly degenerate because of the spatially unresolved and often sparse observations. 

\subsubsection{(Sub-)Millimeter Interferometry}
\label{sec:observations:sub-millimeter-interferometry}

The resolved images in dust continuum emission and molecular line emission do not only directly indicate the disk sizes, but also provide strong constrains on the distribution of dust and gas in circumstellar disks, which cannot be achieved by SED modeling alone. While the infrared excess gave indirect evidence of disks, the gas kinematics inferred from early interferometric observations provided strong evidence that the circumstellar material around T~Tauri stars was in disks with a Keplerian rotation profile \citep{1987ApJ...320..336W,1987ApJ...323..294S}. The first large interferometric survey was carried out by \citet{1996A&A...309..493D} with the IRAM Plateau de Bure Interferometer, in which many disks in Taurus were resolved with with typical angular sizes of 1--2$''$, or 140--280\,AU at a distance of 140\,pc. A substantial improvement was provided by the Sub-millimeter Array (SMA) with high sensitivities and angular resolutions \citep{Andrews:2007hv,Andrews:2007p3380,Andrews:2010eg}. A combination of modeling SEDs  and the high spatial resolution images with SMA significantly improved our understanding of the surface density distribution in disks \citep{Andrews:2009jo,Andrews:2010eg}. These SMA data also directly resolved gaps/holes in transition disks which were already inferred from SED modeling \citep[e.g.,][]{Brown:2009hr,Andrews:2011ic}. The typical hole size from these SMA observations are around several tens AUs, which may be due to the limited angular resolutions. From these high angular resolution observations, the typical disk size is around 100\,AU, and the typical disk-to-host mass ratio is around 0.2--0.6\% \citep{Andrews:2009jo,Andrews:2010eg,Andrews:2013ku}. 

Recently \submm interferometry revealed radial variations in the dust-to-gas ratio in several disks \citep[e.g.,][]{Panic:2009dt,Andrews:2012ii,Isella:2012kf,deGregorioMonsalvo:2013jp,Rosenfeld:2013hv} and resolved multi-wavelength observations indicate also radial variations in the dust properties, see \cref{sec:observations:results:spectral-indices-grain-sizes}.

There are several challenges associated with interpreting the \submm interferometric data. For the dust continuum emission, the primary issue are the unknown optical properties of the dust, which mainly depend on the unknown size distribution and internal structure (i.e. porosity, composition, etc.) of the grains. The observations at several \submm wavelengths may give some constraints on the size distribution of dust grains \citep{Ricci:2010gc}. However, a considerable amount of dust mass could be locked up in larger particles or planetesimals, in which case the measurements of dust mass from \submm observations would only be lower limits. Moreover, for a conversion from the \submm fluxes into disk masses, one needs to know the gas-to-dust ratio, which is expected to change with disk radii and with disk evolution \citep{Takeuchi:2002jf,Birnstiel:2010eq}. In circumstellar disks, the bulk of dominant mass constituent, H$_2$, is unobservable due to the lack of a dipole moment. Other molecular lines, e.g. CO and its isotopes, are used to probe disks. The line emissions are then converted to disk masses assuming abundances for the molecules which depend on the complex chemistry, freeze-out, and optical depth and can vary from region to region in disks.

\subsection{Results}\label{sec:observations:results}

\subsubsection{Disk lifetimes}\label{sec:observations:results:disk-lifetimes}
The disk lifetimes are observationally  constrained by counting in star-forming regions or clusters the numbers of young stars that show disk infrared excess. The resulting drop of disk fraction with the age of the cluster/star-forming region then indicates the average disk lifetime. The early study of the nearby Taurus star-forming region at near-infrared (\SI{2.2}{\um}) wavelength suggest disk lifetimes of $\lesssim$ 3--10 Myr \citep{Strom:1989ig}, which was confirmed later by the study at \SI{10}{\um} \citep{Skrutskie:1990gw}. The first systematic and extensive study of disk lifetimes comes from \citet{Haisch:2001bx}. They studied a sample of 8 star-forming regions/clusters with ages ranging from $\lesssim$ 1--30 \si{Myr} at $L$ band (\SI{3.4}{\um}), and derive an overall disk lifetime of $\sim$6\,Myr. With the advent of the Spitzer Space Telescope, our ability to image star-forming regions/clusters at mid-infrared bands (3.5--24\,\mum) has been dramatically increased. A large sample of star-forming regions/clusters with ages from 1\,Myr to tens of Myrs have been surveyed by various projects with Spitzer \citep[e.g.][]{Allen:2004hk,Megeath:2004dj,Luhman:2007bm,Evans:2009bk,2009ApJS..184...18G,Fang:2009kj,Luhman:2010fz,Megeath:2012cn,Fang:2013br,SiciliaAguilar:2015jz}. The Spitzer data generally confirm the previous results on disk lifetimes \citep{SiciliaAguilar:2006kj,Hernandez:2007dy}. Furthermore, due to the excellent sensitivity and high observing efficiency, the Spitzer data can be used to characterize the timescale of ``transition disks'', which is around 0.1--1\,Myr \citep{Muzerolle:2010es,2011ApJ...732...24C}.

The Spitzer observations are not able to resolve the binary systems with separations $\lesssim$2$''$ even in the nearest star-forming regions like Taurus. Thus the disk lifetimes concluded from these data are actually for both single stars and stellar systems. When correcting for stellar systems, \citet{Kraus:2012dh} find that the fraction of disks surrounding the single stars can be around 80\% in Taurus, while less than 40\% of binary systems with separations $\lesssim$40\,AU have disks. A combination of investigations in other nearby star-forming regions indicates that $\sim$2/3 of close ($<$40\,AU) binary systems  disperse their disks within $\lesssim$1\,Myr \citep{Kraus:2012dh}. Thus, the lifetimes of disks surrounding single stars should be longer than those of close binary systems, which could potentially explain why exoplanet host stars perfer single stars \citep{Roell:2012gp}. Beside in close binary systems, disk evolution can be also affected by other factors. One example are nearby massive stars as found in some massive clusters, e.g. Orion and Pismis\,24, where a type of objects, named ``proplyds'', show tails pointing away from the massive stars. This is interpreted as the outer disks of young stars that are being photoevaporated by ultraviolet radiation from the massive stars \citep{1993ApJ...410..696O,1999ApJ...515..669S,2012A&A...539A.119F}. Spitzer observations have revealed a clear anti-correlation between the frequencies of circumstellar disks and the presence of massive stars in several massive clusters harboring extremely massive stars (typically earlier than O5) \citep[e.g., NGC\,2244, NGC\,6611, Pimis\,24, and etc.][]{2007ApJ...660.1532B,2009A&A...496..453G,2012A&A...539A.119F}. The disks in these massive clusters can be dissipated roughly twice as quickly as in clusters/star-forming regions without extremely massive stars \citep{2012A&A...539A.119F}. The disk lifetimes also show dependence on the density of the cluster environment. In sparse stellar associations disks are dissipated more slowly than those in denser (cluster) environments \citep{2013A&A...549A..15F}. Metallicity could also affect disk evolution in a way which is still controversial \citep{2010ApJ...723L.113Y,2012MNRAS.421...78S}.

There are several issues on disk lifetimes that need to be addressed. The infrared observations can only be useful when the disks still have hot/warm inner regions. Thus, the disk fractions counted from these observations may miss some sources with only cold outer disks, i.e. transition disks. The second is that the disk fractions in star-forming regions/clusters could be also biased to higher values since it is hard to identify the diskless young stars. The age estimates of star-forming regions/clusters are also problematic since there are substantial systematic differences on ages between the different sets of tracks \citep{Hillenbrand:2008vr}. An improved age estimates of young clusters indicate the disk lifetimes could be much longer (10--12\,Myr) than we thought before \citep[][see also \citealp{2014ApJ...793L..34P}]{Bell:2013ch}. 

\subsubsection{Density profiles, disk and hole sizes}
\label{sec:observations:results:density-profiles-disk-and-hole-sizes}

Our knowledge of surface density profile of a circumstellar disk was initiated by the minimum-mass solar nebula (MMSN) model, which was constructed by adding enough light elements to the present planetary bodies to reach solar composition, and then spreading this material out over the regions half-way to their neighboring planets. The resulting  total surface density (dust and gas) can be fitted by a power-law function, $\Sigma\propto r^{-3/2}$ \citep{Weidenschilling:1977kq,Hayashi:1981gg}. A similar model was constructed for extrasolar planets, resulting in a similar slope of around 1.4 \citep{Chiang:2013fh}. The observed features of disks, e.g. the SEDs and the distribution of dust emission, can be generally reproduced  assuming the power-law form ($\Sigg\propto r^{-p}$) for the surface density profile \citep{Beckwith:1990hj,Andrews:2007hv}. The power-law index ($p$) can be only constrained by fitting the resolved images of disks from \submm interferometric observations, which suggest that the $p$ values are typically less than 1.5 \citep{Andrews:2007hv}. The typical disk size from these high spatial resolution observations is around several hundred AU \citep{Andrews:2007hv}. Though succeeding in explaining many observed properties of disks, the power-law form for surface density profiles in disks hardly reconcile the observational fact that disk sizes from CO rotational line emission are much larger than those from dust emission \citep{Pietu:2005hq,Panic:2009dt}. A similarity solution of a simple accretion disk with time-independent viscosity ($\nu$) and $\nu\propto r^{\gamma}$ solve the issue by providing a surface density profile \citep{Lust:1952wr,LyndenBell:1974uq,Hartmann:1998dc,Andrews:2009jo},  
\begin{equation}
\Sigg=\Sigma_\mathrm{g,c}~\left(\frac{r}{r_{\rm c}}\right)^{-\gamma}~\exp\left[-\left(\frac{r}{r_{\rm c}}\right)^{2-\gamma}\right],
\label{eq:sigma_self_similar}
\end{equation}
where $\Sigma_\mathrm{g,c}$ is the normalization parameter, $r_\mathrm{c}$ the characteristic scaling radius, and $\gamma$  is the gradient parameter. When $r\ll r_\mathrm{c}$, $\Sigg$ is approximately a power-law, and when $r\gtrsim r_{\rm c}$, $\Sigg$ starts to be dominated by the exponential factor. The profile has been widely used to reproduce the resolved images of the dust emission or the CO rotational line emission of disks \citep{Andrews:2009jo,Isella:2009bc,Andrews:2011ic,Andrews:2012ii}. When the above exponentially tapered surface density profiles are used in the disk models, physical disk sizes will be less direct than using the power law form, and must be defined explicitly by an intensity threshold. Since $r_\mathrm{c}$ is a characteristic scaling radius beyond which the surface density profiles decrease significantly in an exponential form, its values can probably used to compare the sizes of different disks. From the investigation of a sample of resolved images of disks, $r_\mathrm{c}$ shows a wide range from tens to several hundred AU \citep{Hughes:2008fm,Andrews:2009jo,Andrews:2010eg}. The power-law parameter $\gamma$ was found mostly between 0.5 and 1, much less steep than in the MMSN model. It should be noted that those observations were most sensitive to the disk regions beyond the size of the solar system (on which the MMSN model and the model of \citealp{Chiang:2013fh} is based). It is very well possible, that the inner regions of disks show changes in the dust surface density profile. This is expected theoretically when the particle size distribution transitions from being drift-dominated to being fragmentation dominated, see \cref{sec:dust-disk:radial:densities} and  \citet{Birnstiel:2012ft}. Such a behavior was recently observed for the disk around TW Hya disk by \citet{Menu:2014hg}. 

Advances in sensitivity, however showed that the discrepancy of dust and gas disk sizes remained even if the surface density profile of \cref{eq:sigma_self_similar} was used \citep{Andrews:2012ii,Isella:2012kf,Rosenfeld:2013hv,deGregorioMonsalvo:2013jp}. At the low gas densities in the outer disk, even micron-sized particles can decouple from the gas and start drifting inwards. This drift of small particles can thus leave the outermost regions of the disk devoid of dust, causing steep decreases in the dust-to-gas ratio \citep{Birnstiel:2014cb}. The observed discrepancy between dust and gas disk sizes might thus be a finger-print of the radial drift process.

Another advantage of high spatial resolution images from \submm interferometric observations is to directly resolve gaps or holes in some transition disks. The typical gap/hole sizes from the high angular resolution SMA observations are around ten to  tens of AU \citep{Brown:2009hr,Andrews:2011ic}. However, there are still many transition disks with gap/hole sizes less than 10\,AU, inferred from the SED fitting \citep{Espaillat:2012bl,2013ApJ...769..149K}, which could be resolved with ALMA. The causes for the gaps and cavities are still unknown. Proposed mechanisms include clearing by a unseen companion, photoevaporation, opacity effects, or instabilities. In the recent years, it has become clear, that opacity effects from grain growth can be excluded \citep{Birnstiel:2012jr}, and also photoevaporation models alone can currently not account for most of the gap sizes and accretion rates seen in those objects \citep{Owen:2011fd}, but possibly for a subset of them \citep{Owen:2012cw}. The currently favored mechanism is ``dust trapping'': a bump in the radial gas pressure profile is able to stop radial drift, thus trapping large particles \cite[e.g.,][]{Rice:2006ho,Pinilla:2012ke,Zhu:2012kr}, while only a small fraction of small dust particles are following the gas flow into the inner disk. The cause of such a pressure bump could be either a planet embedded in the disk, or variations in the viscosity profile \citep[e.g.,][]{Kretke:2007jn,Regaly:2012fl}.
   
\subsubsection{Spectral indices, grain sizes}
\label{sec:observations:results:spectral-indices-grain-sizes}

The thermally emitted intensity of the vertically isothermal, face-on dust disk at a distance $r$ from the star is approximately
\begin{equation}
I_\nu(r) = B_\nu(T(r))\cdot\left(1-\exp\left(-\tau_\nu(r)\right)\right),
\end{equation} 
where $B_\nu(T(r))$ is the Planck spectrum at a temperature $T$ and $\tau_\nu(r) \simeq \Sigd\,\kappa_\nu$ is the optical depth ($\nu$ now denotes the frequency). If the optical depth is high, the spectral index is entirely set by the Planck spectrum. If the optical depth is low, we can expand the term in brackets to first order and write
\begin{equation}
I_\nu(r) = B_\nu(T(r))\cdot\kappa_\nu(r)\cdot\Sigd(r).
\end{equation} 

In this optical thin case, the spectral index of the flux at \submm wavelength, \alphamm, defined as $F_{\nu}\propto\nu^{\alphamm}$ depends on the dust opacity and can be used to characterize the dust grain sizes \citep{Beckwith:1991ce,Andrews:2005dj,Ricci:2010gc}. The disk-integrated flux is often roughly approximated as 
\begin{equation}
 F_{\nu}\approx\kappa_\nu\,B_{\nu}(T_{\rm d})\,M_{\rm d}\,D^{-2},  
\end{equation}   
where $T_{\rm d}$ is the temperature of the region dominating the dust emission, $M_{\rm d}$ is the dust mass, and $D$ is the distance from the source. Under the assumption that dust emission at \submm wavelengths is in the Rayleigh-Jeans limit, 
\begin{equation}
 F_{\nu} \approx \kappa_\nu\,2\,\kb\,\nu^2\,T_{\rm d}\,M_{\rm d}\,c^{-2}\,D^{-2},
\end{equation}
where \kb is the Boltzmann constant and $c$ the speed of light. At \submm wavelengths, the dust opacity ($\kappa_\nu$) can be approximated as $\kappa_\nu \propto \nu^{\beta}$, where $\beta$ is is sensitive to dust properties, in particular to the grain size \citep{Draine:2006is,Ricci:2010gc,Testi:2014cj} since large dust particles have smaller $\beta$. The observed flux density can be written as $F_{\nu}\propto\nu^{2+\beta}$. The dust properties in disks can be characterized using the observed spectral indices ($\alphamm$) since $\beta=\alphamm-2$. Regions outside of a few AU are expected to be optically thin, apart from local dust concentrations, such as seen in transition disks. The derived values of $\beta$ from the spectral indices of disks are usually much smaller than for ISM-like dust grains, indicating that dust grains in disk have grown to considerably larger sizes \citep{Beckwith:1991ce,Testi:2003dr,Natta:2004hu,Rodmann:2006kk,Andrews:2007hv,Ricci:2010gc,Ricci:2010bn}. While the \submm wavelengths reveal dust growth in the disk interior, dust growth in the disk atmosphere is also evidenced by the broad and weak silicate 10$\mu m$ emission features which are expected from silicates with larger sizes than those in ISM \citep{vanBoekel:2005fh}.

In recent years, the increasing resolution of \submm interferometers enabled radially resolved measurements of \alphamm, which can put constrains on the radial distribution of grain sizes \citep{Isella:2010ji,Guilloteau:2011ek,Perez:2012ii,Trotta:2013cj}. It was generally found that the spectral index is radially increasing, which indicates larger grains in the inner disk, in agreement with theoretical expectations.

\subsubsection{Recent imaging results}
\label{sec:observations:results:recent-imaging-results-nir-au-and-alma}

The field of protoplanetary disks is currently being revolutionized by observational advances in near-infrared imaging (Subaru, NACO, and the upcoming Sphere and GPI instruments) and in \submm interferometry with ALMA. These observations of disks can provide images with comparable angular resolution (of the order of 0.03$''$), but are sensitive to dust particles of very different sizes ($\mu m$ vs. mm) and temperatures (hot vs. cold). In near-infrared bands, imaging can resolve the inner regions of circumstellar disks with inner working angles of $<0.1''$, sensitive to the $\mu m$-sized dust grains. Observations with Subaru confirm the gaps/holes in some transition disks, which are resolved with \submm interferometric observations \citep{Hashimoto:2012cd,Mayama:2012ez,2014ApJ...783...90T}. It is even more interesting that some transition disks that show  gaps/holes in the \submm interferometric data lack evidence of evolved inner disks in $H$-band polarized intensity images \citep{Muto:2012is,Follette:2013jf}, which may be due to the decoupling between the spatial distributions of the small ($\mu m$) and big (mm) dust grains inside the holes/gaps caused by dust filtration \citep{Rice:2006ho,Dong:2012il,Pinilla:2012ke,Zhu:2012kr}. Toward the transition disk, HD~135344B, the Subaru polarized  $H$-band intensity image reveals the spiral structures in the disk, which may be due to the gravitational perturbation from an embedded planet \citep{Muto:2012is}. Imaging with VLT/NACO revealed gaps, brightness asymmetries, or spiral patterns that might be indicative of planets \citep[e.g.,][]{Quanz:2013ii,Quanz:2013di,Avenhaus:2014hz}. Similar features have recently been imaged in the first results from GPI \citep{Rapson:2015km}, and SPHERE \citep{Benisty:2015da}. 

With its extremely high angular resolution and sensitivity, ALMA has provided many ground breaking findings already. \citet{vanderMarel:2013ky} discovered that the entire emission from large (mm-sized) dust particles, trapped in the outer regions of a transition disk Oph~IRS~48 comes from one side of the disk, a region spanning less than one third of the orbit. A possible explanation for this is the combination of a radial dust trap and an azimuthal asymmetry such as a vortex. This would cause dust to be radially drifting towards the pressure bump where it further concentrates in the azimuthal asymmetry. Such a dust trap in Oph~IRS~48 seems consistent with theoretical predictions \citep{Pinilla:2012ke,Birnstiel:2013es}, and may support the formation of analogues to Kuiper Belt objects in this object. Similar asymmetries in the ring-like dust emission from transition disks have been detected in \citet{Casassus:2013ky} and \citet{Perez:2014fe}. \citet{Casassus:2013ky} also detected faint stream-like features in HCO$^+$ and dust continuum at 345~GHz in the transition disk, HD~142527, which could be explained as funnel flows into the inner disk through  planets \citep{Zhu:2011io}. 

Recently ALMA observations revealed an intriguing image of a circumstellar disk in HL~Tau \citep{ALMAPartnership:2015cg} which is a Class\,I young stellar object, earlier than the typical Class\,II T~Tauri stars. In the disk of HL~Tau, several concentric dust gaps and rings can be clearly identified, possibly sculpted by young planets \citep{Tamayo:2015dt}, which hint that planets could already be formed at very young disk ages. Other proposed explanations include pressure bumps formed due to magnetised turbulence \citep{Johansen:2009jf,Dittrich:2013ix,Flock:2015ks} or condensation fronts \citep{Cuzzi:2004kx,Ros:2013ey,Zhang:2015id,Okuzumi:2015wh}. Whether such sub-structure is a common, yet currently unresolved feature of disks remains to be seen.

\section{Summary}\label{sec:summary}

Small dust particles, starting out with sizes of a micrometer or less, are not only the material out of which planets form, they are also a key ingredient to protoplanetary disks: they are the main source of opacity, thus determining the temperature structure, the evolution of the disk, and ultimately the observational appearance. They also provide the surface area for chemical reactions that produce complex organic molecules, the pre-biotic building blocks of life. Recent years have seen a true revolution in the observational capabilities and with facilities such as ALMA, Sphere, or GPI just getting started, the pace is not going to slow down. Our ability to interpret these observations of disks rests on a good understanding of the interaction of solids and gas and therefore also on the evolution of particle sizes in disks. In this chapter, we reviewed some of the current ideas about how particles grow, how they are transported, and how they eventually make up the building blocks of planets. The key findings are summarized as follows.  
 
Growth and transport processes are driven by interaction between the gas and the dust through drag forces. The drag forces depend on the particle size, hence, they induce collisions between the particles. The collisions can result in a wide variety of outcomes that depend on the collision speed and particle properties. The coarsest classification of collision outcomes would be growth (mass gain of the larger particle), fragmentation/erosion (mass loss of at least one particle, resulting in small fragments), or bouncing collisions (growth neutral, but possibly compaction). The change in size or mass of the particle in turn alters its aerodynamic behavior, hence the size evolution and the global transport of particles cannot be treated separately.

Small grains are well coupled to the gas motion, so they mostly follow the motion of the gas along (e.g., accretion, viscous spreading, turbulent mixing). But as their surface to mass ratio decreases with particle growth, particles start to decouple from the gas flow resulting in substantial velocity differences between dust and gas. This drift motion of the dust relative to the gas tends to move dust towards higher pressure regions, where the dust-gas relative velocity is reduced or vanishes. In a typical model of a protoplanetary disk, this means that particles sediment towards the mid-plane and spiral inwards. The time scale for this orbital decay is quite short, of the order of a few hundred orbits. This motion tends to deplete the surface layers and the outer regions of disks, but regions of high pressure, particularly pressure maximums are able to collect or ``trap'' the dust particles and cause accumulations of dust.

But as particles grow not only do their drift speeds increase, but also the collision they experience happen at higher velocities. This tends to lead to bouncing, fragmentation, or erosion instead of continued growth. Even if particles continued to grow, the increasing drift motion removes them faster than growth is able to resupply them. One way or another, particles are expected to reach only a finite maximum size, typically less than a meter, unless some mechanism is able to circumvent or avoid these issues. Possible pathways to continued growth include the following:
\begin{itemize}
  \item 
 Stochastic effects (``lucky particles'') together with the effect that large particles gain mass by being bombarded by many small grains allow growth of a small number of large bodies. But the time scales for this growth are still longer than the drift motion.
 \item
 Icy particles are expected to be much harder to compress. Their increased surface to mass ratio combined with aerodynamic effects in the inner regions of disks can lead to very short growth time scales. However erosion or sintering might limit this mechanism.
\item
Other solutions to this problem may be the aforementioned particle traps, for example vortices that can efficiently accumulate large amounts of dust. Alternatively local two-fluid instabilities, as discussed in \cref{sec:planetesimal-formation:gravitational:streaming-instability} may lead to gravitationally bound clumps of particles.
\end{itemize}
The latter accumulation mechanisms, however, rely on particle sizes being much larger than the micrometer sized dust inherited from the interstellar medium. Hence, the most likely pathway towards the building blocks of planets seems to be a hybrid approach where particle growth to macroscopic sizes is able to provide particle sizes that can efficiently be accumulated by aerodynamic effects to build up gravitationally bound clumps.

Observations over the last few years have started to provide the crucial input for theoretical models of disk evolution and planetesimal formation. Until recently, the radial drift motion of dust grains was merely a theoretical expectation. But now, observed structures in transition disks, extreme dust over-densities, and different sizes of dust and gas disks seem impossible to explain without it. Radially resolved multi-wavelength observations indicate larger grains being more centrally confined than smaller grains. The seemingly upper limit of the grain size distribution and its radial variations are in reasonable agreement with theoretical models.

Still crucial questions remain unanswered. There is no consistent picture of disk evolution or how the observed morphology of disks can be categorized into one or several evolutionary sequences. The role of various disk evolution mechanisms such as photoevaporation, disk winds, and even the driver of disk accretion remains unknown, let alone their impact on the solids in the disk. Several recent high-resolution observations in both near-infrared and in \submm interferometry have revealed striking morphologies in disks that seem to pose more questions than they answer. Our future understanding of planet formation and disk evolution will crucially depend on larger samples at high resolution and multiple wavelengths that are able to reveal correlations, frequencies, and the full diversity in disk morphology. At the same time, theoretical models will need to evolve to include all relevant physical effects, not just a subset. Furthermore, they need to provide clear predictions that can be tested observationally. With a number of facilities providing a wealth of observations, at resolutions and sensitivities orders of magnitude better than before, the time seems ripe to answer some of the most fundamental questions of disk evolution and the early stages of planet formation.

\begin{acknowledgements}
We like to thank Liubin Pan, Christian Lenz, and the referees, Willy Benz and Alessandro Morbidelli for providing helpful comments that have significantly improved the manuscript.
T.B. acknowledges support from the NASA Origins of Solar Systems grant NNX12AJ04G and from the DFG through grant KL 1469/13-1.
M.F. acknowledges support by the NSFC through grant 11203081.
A.J. is grateful for support from a Starting Grant from the European Research Council (278675-PEBBLE2PLANET) and the Knut and Alice Wallenberg Foundation and the Swedish Research Council (2014-5775).
\end{acknowledgements}

\bibliographystyle{apj}
\bibliography{bibliography}                

\begin{thebibliography}{}
\expandafter\ifx\csname natexlab\endcsname\relax\def\natexlab#1{#1}\fi

\bibitem[{Adams {et~al.}(1987)Adams, Lada, \& Shu}]{Adams:1987gy}
Adams, F.~C., Lada, C.~J., \& Shu, F.~H. 1987, ApJ, 312, 788

\bibitem[{Allen {et~al.}(2004)Allen, Calvet, D'Alessio, Merin, Hartmann,
  Megeath, Gutermuth, Muzerolle, Pipher, Myers, \& Fazio}]{Allen:2004hk}
Allen, L.~E., Calvet, N., D'Alessio, P., {et~al.} 2004, ApJS, 154, 363

\bibitem[{ALMA~Partnership {et~al.}(2015)ALMA~Partnership, Brogan, P{\'e}rez,
  Hunter, Dent, Hales, Hills, Corder, Fomalont, Vlahakis, Asaki, Barkats,
  Hirota, Hodge, Impellizzeri, Kneissl, Liuzzo, Lucas, Marcelino, Matsushita,
  Nakanishi, Phillips, Richards, Toledo, Aladro, Broguiere, Cortes, Cortes,
  Espada, Galarza, Garcia-Appadoo, Guzman-Ramirez, Humphreys, Jung, Kameno,
  Laing, Leon, Marconi, Mignano, Nikolic, Nyman, Radiszcz, Remijan, Rod{\'o}n,
  Sawada, Takahashi, Tilanus, Vila~Vilaro, Watson, Wiklind, Akiyama, Chapillon,
  de~Gregorio-Monsalvo, Di~Francesco, Gueth, Kawamura, Lee, Nguyen~Luong,
  Mangum, Pietu, Sanhueza, Saigo, Takakuwa, Ubach, Van~Kempen, Wootten,
  Castro-Carrizo, Francke, Gallardo, Garcia, Gonzalez, Hill, Kaminski, Kurono,
  Liu, Lopez, Morales, Plarre, Schieven, Testi, Videla, Villard, Andreani,
  Hibbard, \& Tatematsu}]{ALMAPartnership:2015cg}
ALMA~Partnership, {\textasciitilde}., Brogan, C.~L., P{\'e}rez, L.~M., {et~al.}
  2015, ApJL, 808, L3

\bibitem[{Andrews(2015)}]{Andrews:2015cg}
Andrews, S.~M. 2015, Publications of the Astronomical Society of Pacific, 127,
  961

\bibitem[{Andrews {et~al.}(2013)Andrews, Rosenfeld, Kraus, \&
  Wilner}]{Andrews:2013ku}
Andrews, S.~M., Rosenfeld, K.~A., Kraus, A.~L., \& Wilner, D.~J. 2013, ApJ,
  771, 129

\bibitem[{Andrews \& Williams(2005)}]{Andrews:2005dj}
Andrews, S.~M., \& Williams, J.~P. 2005, ApJ, 631, 1134

\bibitem[{Andrews \& Williams(2007{\natexlab{a}})}]{Andrews:2007hv}
---. 2007{\natexlab{a}}, ApJ, 659, 705

\bibitem[{Andrews \& Williams(2007{\natexlab{b}})}]{Andrews:2007p3380}
---. 2007{\natexlab{b}}, ApJ, 671, 1800

\bibitem[{Andrews {et~al.}(2011)Andrews, Wilner, Espaillat, Hughes, Dullemond,
  McClure, Qi, \& Brown}]{Andrews:2011ic}
Andrews, S.~M., Wilner, D.~J., Espaillat, C., {et~al.} 2011, ApJ, 732, 42

\bibitem[{Andrews {et~al.}(2009)Andrews, Wilner, Hughes, Qi, \&
  Dullemond}]{Andrews:2009jo}
Andrews, S.~M., Wilner, D.~J., Hughes, A.~M., Qi, C., \& Dullemond, C.~P. 2009,
  ApJ, 700, 1502

\bibitem[{Andrews {et~al.}(2010)Andrews, Wilner, Hughes, Qi, \&
  Dullemond}]{Andrews:2010eg}
---. 2010, ApJ, 723, 1241

\bibitem[{Andrews {et~al.}(2012)Andrews, Wilner, Hughes, Qi, Rosenfeld,
  {\"O}berg, Birnstiel, Espaillat, Cieza, Williams, Lin, \&
  Ho}]{Andrews:2012ii}
Andrews, S.~M., Wilner, D.~J., Hughes, A.~M., {et~al.} 2012, ApJ, 744, 162

\bibitem[{{Apai} {et~al.}(2005){Apai}, {Pascucci}, {Bouwman}, {Natta},
  {Henning}, \& {Dullemond}}]{2005Sci...310..834A}
{Apai}, D., {Pascucci}, I., {Bouwman}, J., {et~al.} 2005, Science, 310, 834

\bibitem[{Armitage(2011)}]{Armitage:2011fm}
Armitage, P.~J. 2011, ARA{\&}A, 49, 195

\bibitem[{Ataiee {et~al.}(2013)Ataiee, Pinilla, Zsom, Dullemond, Dominik, \&
  Ghanbari}]{Ataiee:2013kx}
Ataiee, S., Pinilla, P., Zsom, A., {et~al.} 2013, A{\&}A, 553, L3

\bibitem[{Avenhaus {et~al.}(2014)Avenhaus, Quanz, Schmid, Meyer, Garufi, Wolf,
  \& Dominik}]{Avenhaus:2014hz}
Avenhaus, H., Quanz, S.~P., Schmid, H.~M., {et~al.} 2014, ApJ, 781, 87

\bibitem[{Bai \& Stone(2010)}]{Bai:2010gm}
Bai, X.-N., \& Stone, J.~M. 2010, ApJ, 722, 1437

\bibitem[{Balbus \& Hawley(1991)}]{Balbus:1991fi}
Balbus, S.~A., \& Hawley, J.~F. 1991, ApJ, 376, 214

\bibitem[{{Balog} {et~al.}(2007){Balog}, {Muzerolle}, {Rieke}, {Su}, {Young},
  \& {Megeath}}]{2007ApJ...660.1532B}
{Balog}, Z., {Muzerolle}, J., {Rieke}, G.~H., {et~al.} 2007, \apj, 660, 1532

\bibitem[{Barge \& Sommeria(1995)}]{Barge:1995vd}
Barge, P., \& Sommeria, J. 1995, A{\&}A, 295, L1

\bibitem[{Beckwith \& Sargent(1991)}]{Beckwith:1991ce}
Beckwith, S. V.~W., \& Sargent, A.~I. 1991, ApJ, 381, 250

\bibitem[{Beckwith {et~al.}(1990)Beckwith, Sargent, Chini, \&
  Guesten}]{Beckwith:1990hj}
Beckwith, S. V.~W., Sargent, A.~I., Chini, R.~S., \& Guesten, R. 1990, AJ, 99,
  924

\bibitem[{Bell {et~al.}(2013)Bell, Naylor, Mayne, Jeffries, \&
  Littlefair}]{Bell:2013ch}
Bell, C. P.~M., Naylor, T., Mayne, N.~J., Jeffries, R.~D., \& Littlefair, S.~P.
  2013, MNRAS, 434, 806

\bibitem[{Benisty {et~al.}(2015)Benisty, Juh{\'a}sz, Boccaletti, Avenhaus,
  Milli, Thalmann, Dominik, Pinilla, Buenzli, Pohl, Beuzit, Birnstiel, de~Boer,
  Bonnefoy, Chauvin, Christiaens, Garufi, Grady, Henning, Huelamo, Isella,
  Langlois, M{\'e}nard, Mouillet, Olofsson, Pantin, Pinte, \&
  Pueyo}]{Benisty:2015da}
Benisty, M., Juh{\'a}sz, A., Boccaletti, A., {et~al.} 2015, A{\&}A, 578, L6

\bibitem[{Benz \& Asphaug(1999)}]{Benz:1999cj}
Benz, W., \& Asphaug, E. 1999, Icarus, 142, 5

\bibitem[{Birnstiel \& Andrews(2014)}]{Birnstiel:2014cb}
Birnstiel, T., \& Andrews, S.~M. 2014, ApJ, 780, 153

\bibitem[{Birnstiel {et~al.}(2012{\natexlab{a}})Birnstiel, Andrews, \&
  Ercolano}]{Birnstiel:2012jr}
Birnstiel, T., Andrews, S.~M., \& Ercolano, B. 2012{\natexlab{a}}, A{\&}A, 544,
  A79

\bibitem[{Birnstiel {et~al.}(2009)Birnstiel, Dullemond, \&
  Brauer}]{Birnstiel:2009hc}
Birnstiel, T., Dullemond, C.~P., \& Brauer, F. 2009, A{\&}A, 503, L5

\bibitem[{Birnstiel {et~al.}(2010)Birnstiel, Dullemond, \&
  Brauer}]{Birnstiel:2010eq}
---. 2010, A{\&}A, 513, A79

\bibitem[{Birnstiel {et~al.}(2013)Birnstiel, Dullemond, \&
  Pinilla}]{Birnstiel:2013es}
Birnstiel, T., Dullemond, C.~P., \& Pinilla, P. 2013, A{\&}A, 550, L8

\bibitem[{Birnstiel {et~al.}(2012{\natexlab{b}})Birnstiel, Klahr, \&
  Ercolano}]{Birnstiel:2012ft}
Birnstiel, T., Klahr, H., \& Ercolano, B. 2012{\natexlab{b}}, A{\&}A, 539, A148

\bibitem[{Birnstiel {et~al.}(2011)Birnstiel, Ormel, \&
  Dullemond}]{Birnstiel:2011ks}
Birnstiel, T., Ormel, C.~W., \& Dullemond, C.~P. 2011, A{\&}A, 525, A11

\bibitem[{Blum \& Wurm(2008)}]{Blum:2008fi}
Blum, J., \& Wurm, G. 2008, ARA{\&}A, 46, 21

\bibitem[{Bockel{\'e}e-Morvan {et~al.}(2002)Bockel{\'e}e-Morvan, Gautier,
  Hersant, Hur{\'e}, \& Robert}]{BockeleeMorvan:2002jx}
Bockel{\'e}e-Morvan, D., Gautier, D., Hersant, F., Hur{\'e}, J.-M., \& Robert,
  F. 2002, A{\&}A, 384, 1107

\bibitem[{Brauer {et~al.}(2008)Brauer, Dullemond, \& Henning}]{Brauer:2008bd}
Brauer, F., Dullemond, C.~P., \& Henning, T. 2008, A{\&}A, 480, 859

\bibitem[{Brauer {et~al.}(2007)Brauer, Dullemond, Johansen, Henning, Klahr, \&
  Natta}]{Brauer:2007dd}
Brauer, F., Dullemond, C.~P., Johansen, A., {et~al.} 2007, A{\&}A, 469, 1169

\bibitem[{Brown {et~al.}(2009)Brown, Blake, Qi, Dullemond, Wilner, \&
  Williams}]{Brown:2009hr}
Brown, J.~M., Blake, G.~A., Qi, C., {et~al.} 2009, ApJ, 704, 496

\bibitem[{Carrera {et~al.}(2015)Carrera, Johansen, \& Davies}]{Carrera:2015hz}
Carrera, D., Johansen, A., \& Davies, M.~B. 2015, A{\&}A, 579, A43

\bibitem[{Casassus {et~al.}(2013)Casassus, Van Der~Plas, M, Dent, Fomalont,
  Hagelberg, Hales, Jord{\'a}n, Mawet, M{\'e}nard, Wootten, Wilner, Hughes,
  Schreiber, Girard, Ercolano, Canovas, Rom{\'a}n, \&
  Salinas}]{Casassus:2013ky}
Casassus, S., Van Der~Plas, G., M, S.~P., {et~al.} 2013, Nature, 493, 191

\bibitem[{Chiang \& Goldreich(1997)}]{Chiang:1997ev}
Chiang, E., \& Goldreich, P. 1997, ApJ, 490, 368

\bibitem[{Chiang \& Laughlin(2013)}]{Chiang:2013fh}
Chiang, E., \& Laughlin, G. 2013, MNRAS, 431, 3444

\bibitem[{{Cieza} {et~al.}(2012){Cieza}, {Schreiber}, {Romero}, {Williams},
  {Rebassa-Mansergas}, \& {Mer{\'{\i}}n}}]{2012ApJ...750..157C}
{Cieza}, L.~A., {Schreiber}, M.~R., {Romero}, G.~A., {et~al.} 2012, \apj, 750,
  157

\bibitem[{{Cieza} {et~al.}(2010){Cieza}, {Schreiber}, {Romero}, {Mora},
  {Merin}, {Swift}, {Orellana}, {Williams}, {Harvey}, \&
  {Evans}}]{2010ApJ...712..925C}
---. 2010, \apj, 712, 925

\bibitem[{{Currie} \& {Sicilia-Aguilar}(2011)}]{2011ApJ...732...24C}
{Currie}, T., \& {Sicilia-Aguilar}, A. 2011, \apj, 732, 24

\bibitem[{Cuzzi {et~al.}(2008)Cuzzi, Hogan, \& Shariff}]{Cuzzi:2008js}
Cuzzi, J.~N., Hogan, R.~C., \& Shariff, K. 2008, ApJ, 687, 1432

\bibitem[{Cuzzi \& Zahnle(2004)}]{Cuzzi:2004kx}
Cuzzi, J.~N., \& Zahnle, K.~J. 2004, ApJ, 614, 490

\bibitem[{de~Gregorio-Monsalvo {et~al.}(2013)de~Gregorio-Monsalvo, M{\'e}nard,
  Dent, Pinte, L{\'o}pez, Klaassen, Hales, Cort{\'e}s, Rawlings, Tachihara,
  Testi, Takahashi, Chapillon, Mathews, Juhasz, Akiyama, Higuchi, Saito, Nyman,
  Phillips, Rod{\'o}n, Corder, \& Van~Kempen}]{deGregorioMonsalvo:2013jp}
de~Gregorio-Monsalvo, I., M{\'e}nard, F., Dent, W., {et~al.} 2013, A{\&}A, 557,
  133

\bibitem[{{Dent} {et~al.}(2013){Dent}, {Thi}, {Kamp}, {Williams}, {Menard},
  {Andrews}, {Ardila}, {Aresu}, {Augereau}, {Navascues}, {Brittain}, {Carmona},
  {Ciardi}, {Danchi}, {Donaldson}, {Duchene}, {Eiroa}, {Fedele}, {Grady}, {de
  Gregorio-Molsalvo}, {Howard}, {Hu{\'e}lamo}, {Krivov}, {Lebreton}, {Liseau},
  {Martin-Zaidi}, {Mathews}, {Meeus}, {Mendigut{\'{\i}}a}, {Montesinos},
  {Morales-Calderon}, {Mora}, {Nomura}, {Pantin}, {Pascucci}, {Phillips},
  {Pinte}, {Podio}, {Ramsay}, {Riaz}, {Riviere-Marichalar}, {Roberge},
  {Sandell}, {Solano}, {Tilling}, {Torrelles}, {Vandenbusche}, {Vicente},
  {White}, \& {Woitke}}]{2013PASP..125..477D}
{Dent}, W.~R.~F., {Thi}, W.~F., {Kamp}, I., {et~al.} 2013, \pasp, 125, 477

\bibitem[{Dittrich {et~al.}(2013)Dittrich, Klahr, \&
  Johansen}]{Dittrich:2013ix}
Dittrich, K., Klahr, H., \& Johansen, A. 2013, ApJ, 763, 117

\bibitem[{Dohnanyi(1969)}]{Dohnanyi:1969dw}
Dohnanyi, J.~S. 1969, J. Geophys. Res., 74, 2531

\bibitem[{Dominik \& Dullemond(2011)}]{Dominik:2011ih}
Dominik, C., \& Dullemond, C.~P. 2011, A{\&}A, 531, A101

\bibitem[{Dong {et~al.}(2012)Dong, Rafikov, Zhu, Hartmann, Whitney, Brandt,
  Muto, Hashimoto, Grady, Follette, Kuzuhara, Tanii, Itoh, Thalmann,
  Wisniewski, Mayama, Janson, Abe, Brandner, Carson, Egner, Feldt, Goto, Guyon,
  Hayano, Hayashi, Hayashi, Henning, Hodapp, Honda, Inutsuka, Ishii, Iye,
  Kandori, Knapp, Kudo, Kusakabe, Matsuo, McElwain, Miyama, Morino,
  Moro-Martin, Nishimura, Pyo, Suto, Suzuki, Takami, Takato, Terada, Tomono,
  Turner, Watanabe, Yamada, Takami, Usuda, \& Tamura}]{Dong:2012il}
Dong, R., Rafikov, R., Zhu, Z., {et~al.} 2012, ApJ, 750, 161

\bibitem[{Draine(2006)}]{Draine:2006is}
Draine, B.~T. 2006, ApJ, 636, 1114

\bibitem[{Dr{\k{a}}{\.{z}}kowska \& Dullemond(2014)}]{Drazkowska:2014dj}
Dr{\k{a}}{\.{z}}kowska, J., \& Dullemond, C.~P. 2014, A{\&}A, 572, A78

\bibitem[{Dr{\k{a}}{\.{z}}kowska {et~al.}(2014)Dr{\k{a}}{\.{z}}kowska,
  Windmark, \& Dullemond}]{Drazkowska:2014hp}
Dr{\k{a}}{\.{z}}kowska, J., Windmark, F., \& Dullemond, C.~P. 2014, A{\&}A,
  567, 38

\bibitem[{Dubrulle {et~al.}(1995)Dubrulle, Morfill, \&
  Sterzik}]{Dubrulle:1995jn}
Dubrulle, B., Morfill, G.~E., \& Sterzik, M.~F. 1995, Icarus, 114, 237

\bibitem[{Dullemond(2012)}]{Dullemond:2012vq}
Dullemond, C. 2012, ASCL, -1, 02015

\bibitem[{Dullemond \& Dominik(2004)}]{Dullemond:2004iy}
Dullemond, C.~P., \& Dominik, C. 2004, A{\&}A, 417, 159

\bibitem[{Dullemond \& Dominik(2005)}]{Dullemond:2005hf}
---. 2005, A{\&}A, 434, 971

\bibitem[{{Dutrey} {et~al.}(1996){Dutrey}, {Guilloteau}, {Duvert}, {Prato},
  {Simon}, {Schuster}, \& {Menard}}]{1996A&A...309..493D}
{Dutrey}, A., {Guilloteau}, S., {Duvert}, G., {et~al.} 1996, \aap, 309, 493

\bibitem[{Espaillat {et~al.}(2012)Espaillat, Ingleby, Hern{\'a}ndez, Furlan,
  D'Alessio, Calvet, Andrews, Muzerolle, Qi, \& Wilner}]{Espaillat:2012bl}
Espaillat, C., Ingleby, L., Hern{\'a}ndez, J., {et~al.} 2012, ApJ, 747, 103

\bibitem[{Espaillat {et~al.}(2014)Espaillat, Muzerolle, Najita, Andrews, Zhu,
  Calvet, Kraus, Hashimoto, Kraus, \& D'Alessio}]{Espaillat:2014hh}
Espaillat, C., Muzerolle, J., Najita, J., {et~al.} 2014, PPVI, 497

\bibitem[{Evans {et~al.}(2009)Evans, Dunham, J{\o}rgensen, Enoch, Mer{\'\i}n,
  van Dishoeck, Alcal{\'a}, Myers, Stapelfeldt, Huard, Allen, Harvey, van
  Kempen, Blake, Koerner, Mundy, Padgett, \& Sargent}]{Evans:2009bk}
Evans, N.~J., Dunham, M.~M., J{\o}rgensen, J.~K., {et~al.} 2009, ApJS, 181, 321

\bibitem[{Fang {et~al.}(2013)Fang, Kim, van Boekel, Sicilia-Aguilar, Henning,
  \& Flaherty}]{Fang:2013br}
Fang, M., Kim, J.~S., van Boekel, R., {et~al.} 2013, ApJS, 207, 5

\bibitem[{{Fang} {et~al.}(2013){Fang}, {van Boekel}, {Bouwman}, {Henning},
  {Lawson}, \& {Sicilia-Aguilar}}]{2013A&A...549A..15F}
{Fang}, M., {van Boekel}, R., {Bouwman}, J., {et~al.} 2013, \aap, 549, A15

\bibitem[{Fang {et~al.}(2009)Fang, van Boekel, Wang, Carmona, Sicilia-Aguilar,
  \& Henning}]{Fang:2009kj}
Fang, M., van Boekel, R., Wang, W., {et~al.} 2009, A{\&}A, 504, 461

\bibitem[{{Fang} {et~al.}(2012){Fang}, {van Boekel}, {King}, {Henning},
  {Bouwman}, {Doi}, {Okamoto}, {Roccatagliata}, \&
  {Sicilia-Aguilar}}]{2012A&A...539A.119F}
{Fang}, M., {van Boekel}, R., {King}, R.~R., {et~al.} 2012, \aap, 539, A119

\bibitem[{Flock {et~al.}(2015)Flock, Ruge, Dzyurkevich, Henning, Klahr, \&
  Wolf}]{Flock:2015ks}
Flock, M., Ruge, J.~P., Dzyurkevich, N., {et~al.} 2015, A{\&}A, 574, A68

\bibitem[{Follette {et~al.}(2013)Follette, Tamura, Hashimoto, Whitney, Grady,
  Close, Andrews, Kwon, Wisniewski, Brandt, Mayama, Kandori, Dong, Abe,
  Brandner, Carson, Currie, Egner, Feldt, Goto, Guyon, Hayano, Hayashi,
  Hayashi, Henning, Hodapp, Ishii, Iye, Janson, Knapp, Kudo, Kusakabe,
  Kuzuhara, McElwain, Matsuo, Miyama, Morino, Moro-Martin, Nishimura, Pyo,
  Serabyn, Suto, Suzuki, Takami, Takato, Terada, Thalmann, Tomono, Turner,
  Watanabe, Yamada, Takami, \& Usuda}]{Follette:2013jf}
Follette, K.~B., Tamura, M., Hashimoto, J., {et~al.} 2013, ApJ, 767, 10

\bibitem[{Fromang \& Nelson(2009)}]{Fromang:2009ja}
Fromang, S., \& Nelson, R.~P. 2009, A{\&}A, 496, 597

\bibitem[{Fromang \& Papaloizou(2006)}]{Fromang:2006hk}
Fromang, S., \& Papaloizou, J. 2006, A{\&}A, 452, 751

\bibitem[{Fu {et~al.}(2014)Fu, Li, Lubow, Li, \& Liang}]{Fu:2014gk}
Fu, W., Li, H., Lubow, S., Li, S., \& Liang, E. 2014, ApJL, 795, L39

\bibitem[{Furlan {et~al.}(2006)Furlan, Hartmann, Calvet, D'Alessio,
  Franco-Hern{\'a}ndez, Forrest, Watson, Uchida, Sargent, Green, Keller, \&
  Herter}]{Furlan:2006bb}
Furlan, E., Hartmann, L., Calvet, N., {et~al.} 2006, ApJS, 165, 568

\bibitem[{Garaud(2007)}]{Garaud:2007ks}
Garaud, P. 2007, ApJ, 671, 2091

\bibitem[{Garaud {et~al.}(2013)Garaud, Meru, Galvagni, \&
  Olczak}]{Garaud:2013ja}
Garaud, P., Meru, F., Galvagni, M., \& Olczak, C. 2013, ApJ, 764, 146

\bibitem[{{Guarcello} {et~al.}(2009){Guarcello}, {Micela}, {Damiani}, {Peres},
  {Prisinzano}, \& {Sciortino}}]{2009A&A...496..453G}
{Guarcello}, M.~G., {Micela}, G., {Damiani}, F., {et~al.} 2009, \aap, 496, 453

\bibitem[{Guilloteau {et~al.}(2011)Guilloteau, Dutrey, Pi{\'e}tu, \&
  Boehler}]{Guilloteau:2011ek}
Guilloteau, S., Dutrey, A., Pi{\'e}tu, V., \& Boehler, Y. 2011, A{\&}A, 529,
  105

\bibitem[{Gundlach \& Blum(2015)}]{Gundlach:2015iu}
Gundlach, B., \& Blum, J. 2015, ApJ, 798, 34

\bibitem[{Gundlach {et~al.}(2011)Gundlach, Kilias, Beitz, \&
  Blum}]{Gundlach:2011jj}
Gundlach, B., Kilias, S., Beitz, E., \& Blum, J. 2011, Icarus, 214, 717

\bibitem[{{Gutermuth} {et~al.}(2009){Gutermuth}, {Megeath}, {Myers}, {Allen},
  {Pipher}, \& {Fazio}}]{2009ApJS..184...18G}
{Gutermuth}, R.~A., {Megeath}, S.~T., {Myers}, P.~C., {et~al.} 2009, \apjs,
  184, 18

\bibitem[{G{\"u}ttler {et~al.}(2010)G{\"u}ttler, Blum, Zsom, Ormel, \&
  Dullemond}]{Guttler:2010ia}
G{\"u}ttler, C., Blum, J., Zsom, A., Ormel, C.~W., \& Dullemond, C.~P. 2010,
  A{\&}A, 513, 56

\bibitem[{Haisch {et~al.}(2001)Haisch, Lada, \& Lada}]{Haisch:2001bx}
Haisch, K.~E., Lada, E.~A., \& Lada, C.~J. 2001, ApJ, 553, L153

\bibitem[{Hartmann {et~al.}(1998)Hartmann, Calvet, Gullbring, \&
  D'Alessio}]{Hartmann:1998dc}
Hartmann, L., Calvet, N., Gullbring, E., \& D'Alessio, P. 1998, ApJ, 495, 385

\bibitem[{{Harvey} {et~al.}(2012){Harvey}, {Henning}, {Liu}, {M{\'e}nard},
  {Pinte}, {Wolf}, {Cieza}, {Evans}, \& {Pascucci}}]{2012ApJ...755...67H}
{Harvey}, P.~M., {Henning}, T., {Liu}, Y., {et~al.} 2012, \apj, 755, 67

\bibitem[{Hashimoto {et~al.}(2012)Hashimoto, Dong, Kudo, Honda, McClure, Zhu,
  Muto, Wisniewski, Abe, Brandner, Brandt, Carson, Egner, Feldt, Fukagawa,
  Goto, Grady, Guyon, Hayano, Hayashi, Hayashi, Henning, Hodapp, Ishii, Iye,
  Janson, Kandori, Knapp, Kusakabe, Kuzuhara, Kwon, Matsuo, Mayama, McElwain,
  Miyama, Morino, Moro-Martin, Nishimura, Pyo, Serabyn, Suenaga, Suto, Suzuki,
  Takahashi, Takami, Takato, Terada, Thalmann, Tomono, Turner, Watanabe,
  Yamada, Takami, Usuda, \& Tamura}]{Hashimoto:2012cd}
Hashimoto, J., Dong, R., Kudo, T., {et~al.} 2012, ApJL, 758, L19

\bibitem[{Hayashi(1981)}]{Hayashi:1981gg}
Hayashi, C. 1981, Prog. Theor. Phys. Suppl., 70, 35

\bibitem[{Heim {et~al.}(1999)Heim, Blum, Preuss, \& Butt}]{Heim:1999hn}
Heim, L.-O., Blum, J., Preuss, M., \& Butt, H.-J. 1999, Phys. Rev. Lett., 83,
  3328

\bibitem[{Hern{\'a}ndez {et~al.}(2007)Hern{\'a}ndez, Hartmann, Megeath,
  Gutermuth, Muzerolle, Calvet, Vivas, Brice{\~n}o, Allen, Stauffer, Young, \&
  Fazio}]{Hernandez:2007dy}
Hern{\'a}ndez, J., Hartmann, L., Megeath, T., {et~al.} 2007, ApJ, 662, 1067

\bibitem[{Hillenbrand {et~al.}(2008)Hillenbrand, Bauermeister, \&
  White}]{Hillenbrand:2008vr}
Hillenbrand, L.~A., Bauermeister, A., \& White, R.~J. 2008, 14th Cambridge
  Workshop on Cool Stars, 384, 200

\bibitem[{Hsieh \& Gu(2012)}]{Hsieh:2012en}
Hsieh, H.-F., \& Gu, P.-G. 2012, ApJ, 760, 119

\bibitem[{Hughes {et~al.}(2008)Hughes, Wilner, Qi, \&
  Hogerheijde}]{Hughes:2008fm}
Hughes, A.~M., Wilner, D.~J., Qi, C., \& Hogerheijde, M.~R. 2008, ApJ, 678,
  1119

\bibitem[{Isella {et~al.}(2009)Isella, Carpenter, \& Sargent}]{Isella:2009bc}
Isella, A., Carpenter, J.~M., \& Sargent, A.~I. 2009, ApJ, 701, 260

\bibitem[{Isella {et~al.}(2010)Isella, Carpenter, \& Sargent}]{Isella:2010ji}
---. 2010, ApJ, 714, 1746

\bibitem[{Isella {et~al.}(2012)Isella, P{\'e}rez, \& Carpenter}]{Isella:2012kf}
Isella, A., P{\'e}rez, L.~M., \& Carpenter, J.~M. 2012, ApJ, 747, 136

\bibitem[{Johansen {et~al.}(2004)Johansen, Andersen, \&
  Brandenburg}]{Johansen:2004jp}
Johansen, A., Andersen, A.~C., \& Brandenburg, A. 2004, A{\&}A, 417, 361

\bibitem[{Johansen {et~al.}(2014)Johansen, Blum, Tanaka, Ormel, Bizzarro, \&
  Rickman}]{Johansen:2014hs}
Johansen, A., Blum, J., Tanaka, H., {et~al.} 2014, PPVI, 547

\bibitem[{Johansen {et~al.}(2015{\natexlab{a}})Johansen, Jacquet, Cuzzi,
  Morbidelli, \& Gounelle}]{Johansen:2015tc}
Johansen, A., Jacquet, E., Cuzzi, J.~N., Morbidelli, A., \& Gounelle, M.
  2015{\natexlab{a}}, arXiv, 1505, 2941

\bibitem[{Johansen \& Klahr(2005)}]{Johansen:2005eq}
Johansen, A., \& Klahr, H. 2005, ApJ, 634, 1353

\bibitem[{Johansen {et~al.}(2011)Johansen, Klahr, \& Henning}]{Johansen:2011jo}
Johansen, A., Klahr, H., \& Henning, T. 2011, A{\&}A, 529, 62

\bibitem[{Johansen {et~al.}(2015{\natexlab{b}})Johansen, Mac~Low, Lacerda, \&
  Bizzarro}]{Johansen:2015he}
Johansen, A., Mac~Low, M.-M., Lacerda, P., \& Bizzarro, M. 2015{\natexlab{b}},
  Science Advances, 1, 1500109

\bibitem[{Johansen {et~al.}(2007)Johansen, Oishi, Mac~Low, Klahr, Henning, \&
  Youdin}]{Johansen:2007cl}
Johansen, A., Oishi, J.~S., Mac~Low, M.-M., {et~al.} 2007, Nature, 448, 1022

\bibitem[{Johansen {et~al.}(2009)Johansen, Youdin, \& Klahr}]{Johansen:2009jf}
Johansen, A., Youdin, A., \& Klahr, H. 2009, ApJ, 697, 1269

\bibitem[{Johansen {et~al.}(2012)Johansen, Youdin, \&
  Lithwick}]{Johansen:2012kr}
Johansen, A., Youdin, A.~N., \& Lithwick, Y. 2012, A{\&}A, 537, A125

\bibitem[{Kataoka {et~al.}(2013)Kataoka, Tanaka, Okuzumi, \&
  Wada}]{Kataoka:2013kx}
Kataoka, A., Tanaka, H., Okuzumi, S., \& Wada, K. 2013, A{\&}A, 557, L4

\bibitem[{{Kenyon} \& {Hartmann}(1987)}]{1987ApJ...323..714K}
{Kenyon}, S.~J., \& {Hartmann}, L. 1987, \apj, 323, 714

\bibitem[{{Kim} {et~al.}(2013){Kim}, {Watson}, {Manoj}, {Forrest}, {Najita},
  {Furlan}, {Sargent}, {Espaillat}, {Muzerolle}, {Megeath}, {Calvet}, {Green},
  \& {Arnold}}]{2013ApJ...769..149K}
{Kim}, K.~H., {Watson}, D.~M., {Manoj}, P., {et~al.} 2013, \apj, 769, 149

\bibitem[{Klahr \& Bodenheimer(2006)}]{Klahr:2006iv}
Klahr, H., \& Bodenheimer, P. 2006, ApJ, 639, 432

\bibitem[{Klahr \& Bodenheimer(2003)}]{Klahr:2003kq}
Klahr, H.~H., \& Bodenheimer, P. 2003, ApJ, 582, 869

\bibitem[{Klahr \& Henning(1997)}]{Klahr:1997cp}
Klahr, H.~H., \& Henning, T. 1997, Icarus, 128, 213

\bibitem[{Kothe {et~al.}(2010)Kothe, G{\"u}ttler, \& Blum}]{Kothe:2010bd}
Kothe, S., G{\"u}ttler, C., \& Blum, J. 2010, ApJ, 725, 1242

\bibitem[{Kraus {et~al.}(2012)Kraus, Ireland, Hillenbrand, \&
  Martinache}]{Kraus:2012dh}
Kraus, A.~L., Ireland, M.~J., Hillenbrand, L.~A., \& Martinache, F. 2012, ApJ,
  745, 19

\bibitem[{Krauss \& Wurm(2005)}]{Krauss:2005gq}
Krauss, O., \& Wurm, G. 2005, ApJ, 630, 1088

\bibitem[{Kretke \& Lin(2007)}]{Kretke:2007jn}
Kretke, K.~A., \& Lin, D. N.~C. 2007, ApJ, 664, L55

\bibitem[{Kretke {et~al.}(2009)Kretke, Lin, Garaud, \& Turner}]{Kretke:2009ex}
Kretke, K.~A., Lin, D. N.~C., Garaud, P., \& Turner, N.~J. 2009, ApJ, 690, 407

\bibitem[{Krijt {et~al.}(2015)Krijt, Ormel, Dominik, \& Tielens}]{Krijt:2015bu}
Krijt, S., Ormel, C.~W., Dominik, C., \& Tielens, A. G. G.~M. 2015, A{\&}A,
  574, A83

\bibitem[{Krijt {et~al.}(2016)Krijt, Ormel, Dominik, \& Tielens}]{Krijt:2016jm}
---. 2016, A{\&}A, 586, A20

\bibitem[{Laibe {et~al.}(2008)Laibe, Gonzalez, Fouchet, \&
  Maddison}]{Laibe:2008bc}
Laibe, G., Gonzalez, J.-F., Fouchet, L., \& Maddison, S.~T. 2008, A{\&}A, 487,
  265

\bibitem[{Lambrechts \& Johansen(2012)}]{Lambrechts:2012gr}
Lambrechts, M., \& Johansen, A. 2012, A{\&}A, 544, A32

\bibitem[{Lambrechts \& Johansen(2014)}]{Lambrechts:2014iq}
---. 2014, A{\&}A, 572, A107

\bibitem[{Lesur \& Papaloizou(2010)}]{Lesur:2010jn}
Lesur, G., \& Papaloizou, J. C.~B. 2010, A{\&}A, 513, 60

\bibitem[{Luhman(2007)}]{Luhman:2007bm}
Luhman, K.~L. 2007, ApJS, 173, 104

\bibitem[{Luhman {et~al.}(2010)Luhman, Allen, Espaillat, Hartmann, \&
  Calvet}]{Luhman:2010fz}
Luhman, K.~L., Allen, P.~R., Espaillat, C., Hartmann, L., \& Calvet, N. 2010,
  ApJS, 186, 111

\bibitem[{L{\"u}st(1952)}]{Lust:1952wr}
L{\"u}st, R. 1952, Zeitschrift Naturforschung Teil A, 7, 87

\bibitem[{Lynden-Bell \& Pringle(1974)}]{LyndenBell:1974uq}
Lynden-Bell, D., \& Pringle, J.~E. 1974, MNRAS, 168, 603

\bibitem[{Lyra {et~al.}(2008)Lyra, Johansen, Klahr, \& Piskunov}]{Lyra:2008gu}
Lyra, W., Johansen, A., Klahr, H., \& Piskunov, N. 2008, A{\&}A, 479, 883

\bibitem[{Lyra {et~al.}(2009)Lyra, Johansen, Klahr, \& Piskunov}]{Lyra:2009ga}
---. 2009, A{\&}A, 493, 1125

\bibitem[{Lyra \& Lin(2013)}]{Lyra:2013ix}
Lyra, W., \& Lin, M.-K. 2013, ApJ, 775, 17

\bibitem[{Marsh \& Mahoney(1993)}]{Marsh:1993ip}
Marsh, K.~A., \& Mahoney, M.~J. 1993, ApJ, 405, L71

\bibitem[{Mathis {et~al.}(1977)Mathis, Rumpl, \& Nordsieck}]{Mathis:1977hp}
Mathis, J.~S., Rumpl, W., \& Nordsieck, K.~H. 1977, ApJ, 217, 425

\bibitem[{Mayama {et~al.}(2012)Mayama, Hashimoto, Muto, Tsukagoshi, Kusakabe,
  Kuzuhara, Takahashi, Kudo, Dong, Fukagawa, Takami, Momose, Wisniewski,
  Follette, Abe, Akiyama, Brandner, Brandt, Carson, Egner, Feldt, Goto, Grady,
  Guyon, Hayano, Hayashi, Hayashi, Henning, Hodapp, Ishii, Iye, Janson,
  Kandori, Kwon, Knapp, Matsuo, McElwain, Miyama, Morino, Moro-Martin,
  Nishimura, Pyo, Serabyn, Suto, Suzuki, Takato, Terada, Thalmann, Tomono,
  Turner, Watanabe, Yamada, Takami, Usuda, \& Tamura}]{Mayama:2012ez}
Mayama, S., Hashimoto, J., Muto, T., {et~al.} 2012, ApJL, 760, L26

\bibitem[{{Meeus} {et~al.}(2012){Meeus}, {Montesinos}, {Mendigut{\'{\i}}a},
  {Kamp}, {Thi}, {Eiroa}, {Grady}, {Mathews}, {Sandell}, {Martin-Za{\"i}di},
  {Brittain}, {Dent}, {Howard}, {M{\'e}nard}, {Pinte}, {Roberge},
  {Vandenbussche}, \& {Williams}}]{2012A&A...544A..78M}
{Meeus}, G., {Montesinos}, B., {Mendigut{\'{\i}}a}, I., {et~al.} 2012, \aap,
  544, A78

\bibitem[{Megeath {et~al.}(2004)Megeath, Allen, Gutermuth, Pipher, Myers,
  Calvet, Hartmann, Muzerolle, \& Fazio}]{Megeath:2004dj}
Megeath, S.~T., Allen, L.~E., Gutermuth, R.~A., {et~al.} 2004, ApJS, 154, 367

\bibitem[{Megeath {et~al.}(2012)Megeath, Gutermuth, Muzerolle, Kryukova,
  Flaherty, Hora, Allen, Hartmann, Myers, Pipher, Stauffer, Young, \&
  Fazio}]{Megeath:2012cn}
Megeath, S.~T., Gutermuth, R., Muzerolle, J., {et~al.} 2012, AJ, 144, 192

\bibitem[{Meheut {et~al.}(2012)Meheut, Meliani, Varniere, \&
  Benz}]{Meheut:2012di}
Meheut, H., Meliani, Z., Varniere, P., \& Benz, W. 2012, A{\&}A, 545, 134

\bibitem[{Mendoza~V(1966)}]{MendozaV:1966de}
Mendoza~V, E.~E. 1966, ApJ, 143, 1010

\bibitem[{Mendoza~V(1968)}]{MendozaV:1968gp}
---. 1968, ApJ, 151, 977

\bibitem[{Menu {et~al.}(2014)Menu, van Boekel, Henning, Chandler, Linz,
  Benisty, Lacour, Min, Waelkens, Andrews, Calvet, Carpenter, Corder, Deller,
  Greaves, Harris, Isella, Kwon, Lazio, Le~Bouquin, M{\'e}nard, Mundy,
  P{\'e}rez, Ricci, Sargent, Storm, Testi, \& Wilner}]{Menu:2014hg}
Menu, J., van Boekel, R., Henning, T., {et~al.} 2014, A{\&}A, 564, 93

\bibitem[{Muto {et~al.}(2012)Muto, Grady, Hashimoto, Fukagawa, Hornbeck, Sitko,
  Russell, Werren, Cur{\'e}, Currie, Ohashi, Okamoto, Momose, Honda, Inutsuka,
  Takeuchi, Dong, Abe, Brandner, Brandt, Carson, Egner, Feldt, Fukue, Goto,
  Guyon, Hayano, Hayashi, Hayashi, Henning, Hodapp, Ishii, Iye, Janson,
  Kandori, Knapp, Kudo, Kusakabe, Kuzuhara, Matsuo, Mayama, McElwain, Miyama,
  Morino, Moro-Martin, Nishimura, Pyo, Serabyn, Suto, Suzuki, Takami, Takato,
  Terada, Thalmann, Tomono, Turner, Watanabe, Wisniewski, Yamada, Takami,
  Usuda, \& Tamura}]{Muto:2012is}
Muto, T., Grady, C.~A., Hashimoto, J., {et~al.} 2012, ApJL, 748, L22

\bibitem[{Muzerolle {et~al.}(2010)Muzerolle, Allen, Megeath, Hern{\'a}ndez, \&
  Gutermuth}]{Muzerolle:2010es}
Muzerolle, J., Allen, L.~E., Megeath, S.~T., Hern{\'a}ndez, J., \& Gutermuth,
  R.~A. 2010, ApJ, 708, 1107

\bibitem[{{Myers} {et~al.}(1987){Myers}, {Fuller}, {Mathieu}, {Beichman},
  {Benson}, {Schild}, \& {Emerson}}]{1987ApJ...319..340M}
{Myers}, P.~C., {Fuller}, G.~A., {Mathieu}, R.~D., {et~al.} 1987, \apj, 319,
  340

\bibitem[{Nakagawa {et~al.}(1986)Nakagawa, Sekiya, \&
  Hayashi}]{Nakagawa:1986cu}
Nakagawa, Y., Sekiya, M., \& Hayashi, C. 1986, Icarus, 67, 375

\bibitem[{Natta {et~al.}(2004)Natta, Testi, Neri, Shepherd, \&
  Wilner}]{Natta:2004hu}
Natta, A., Testi, L., Neri, R., Shepherd, D.~S., \& Wilner, D.~J. 2004, A{\&}A,
  416, 179

\bibitem[{Nelson {et~al.}(2013)Nelson, Gressel, \& Umurhan}]{Nelson:2013cd}
Nelson, R.~P., Gressel, O., \& Umurhan, O.~M. 2013, MNRAS, 435, 2610

\bibitem[{Nesvorn{\'{y}} {et~al.}(2010)Nesvorn{\'{y}}, Youdin, \&
  Richardson}]{Nesvorny:2010da}
Nesvorn{\'{y}}, D., Youdin, A.~N., \& Richardson, D.~C. 2010, AJ, 140, 785

\bibitem[{Noll {et~al.}(2008)Noll, Grundy, Stephens, Levison, \&
  Kern}]{Noll:2008bj}
Noll, K.~S., Grundy, W.~M., Stephens, D.~C., Levison, H.~F., \& Kern, S.~D.
  2008, Icarus, 194, 758

\bibitem[{{O'dell} {et~al.}(1993){O'dell}, {Wen}, \&
  {Hu}}]{1993ApJ...410..696O}
{O'dell}, C.~R., {Wen}, Z., \& {Hu}, X. 1993, \apj, 410, 696

\bibitem[{Okuzumi {et~al.}(2015)Okuzumi, Momose, Sirono, Kobayashi, \&
  Tanaka}]{Okuzumi:2015wh}
Okuzumi, S., Momose, M., Sirono, S.-I., Kobayashi, H., \& Tanaka, H. 2015,
  eprint arXiv:1510.03556, 1510.03556

\bibitem[{Okuzumi {et~al.}(2012)Okuzumi, Tanaka, Kobayashi, \&
  Wada}]{Okuzumi:2012kd}
Okuzumi, S., Tanaka, H., Kobayashi, H., \& Wada, K. 2012, ApJ, 752, 106

\bibitem[{Okuzumi {et~al.}(2009)Okuzumi, Tanaka, \& Sakagami}]{Okuzumi:2009hf}
Okuzumi, S., Tanaka, H., \& Sakagami, M. 2009, ApJ, 707, 1247

\bibitem[{Ormel \& Cuzzi(2007)}]{Ormel:2007bl}
Ormel, C.~W., \& Cuzzi, J.~N. 2007, A{\&}A, 466, 413

\bibitem[{Ormel {et~al.}(2008)Ormel, Cuzzi, \& Tielens}]{Ormel:2008bn}
Ormel, C.~W., Cuzzi, J.~N., \& Tielens, A. G. G.~M. 2008, ApJ, 679, 1588

\bibitem[{Ormel \& Klahr(2010)}]{Ormel:2010ii}
Ormel, C.~W., \& Klahr, H.~H. 2010, A{\&}A, 520, A43

\bibitem[{Ormel {et~al.}(2009)Ormel, Paszun, Dominik, \&
  Tielens}]{Ormel:2009dq}
Ormel, C.~W., Paszun, D., Dominik, C., \& Tielens, A. G. G.~M. 2009, A{\&}A,
  502, 845

\bibitem[{Ormel \& Spaans(2008)}]{Ormel:2008bp}
Ormel, C.~W., \& Spaans, M. 2008, ApJ, 684, 1291

\bibitem[{Ormel {et~al.}(2007)Ormel, Spaans, \& Tielens}]{Ormel:2007bh}
Ormel, C.~W., Spaans, M., \& Tielens, A. G. G.~M. 2007, A{\&}A, 461, 215

\bibitem[{Owen \& Clarke(2012)}]{Owen:2012cw}
Owen, J.~E., \& Clarke, C.~J. 2012, MNRAS, 426, L96

\bibitem[{Owen {et~al.}(2011)Owen, Ercolano, \& Clarke}]{Owen:2011fd}
Owen, J.~E., Ercolano, B., \& Clarke, C.~J. 2011, MNRAS, 412, 13

\bibitem[{Paardekooper \& Mellema(2004)}]{Paardekooper:2004bf}
Paardekooper, S.-J., \& Mellema, G. 2004, A{\&}A, 425, L9

\bibitem[{Pan \& Padoan(2010)}]{Pan:2010hi}
Pan, L., \& Padoan, P. 2010, Journal of Fluid Mechanics, 661, 73

\bibitem[{Pan {et~al.}(2014)Pan, Padoan, \& Scalo}]{Pan:2014ii}
Pan, L., Padoan, P., \& Scalo, J. 2014, ApJ, 791, 48

\bibitem[{Pani{\'c} {et~al.}(2009)Pani{\'c}, Hogerheijde, Wilner, \&
  Qi}]{Panic:2009dt}
Pani{\'c}, O., Hogerheijde, M.~R., Wilner, D., \& Qi, C. 2009, A{\&}A, 501, 269

\bibitem[{Paraskov {et~al.}(2007)Paraskov, Wurm, \& Krauss}]{Paraskov:2007hm}
Paraskov, G.~B., Wurm, G., \& Krauss, O. 2007, Icarus, 191, 779

\bibitem[{Pavlyuchenkov \& Dullemond(2007)}]{Pavlyuchenkov:2007gw}
Pavlyuchenkov, Y., \& Dullemond, C.~P. 2007, A{\&}A, 471, 833

\bibitem[{P{\'e}rez {et~al.}(2014)P{\'e}rez, Isella, Carpenter, \&
  Chandler}]{Perez:2014fe}
P{\'e}rez, L.~M., Isella, A., Carpenter, J.~M., \& Chandler, C.~J. 2014, ApJL,
  783, L13

\bibitem[{P{\'e}rez {et~al.}(2012)P{\'e}rez, Carpenter, Chandler, Isella,
  Andrews, Ricci, Calvet, Corder, Deller, Dullemond, Greaves, Harris, Henning,
  Kwon, Lazio, Linz, Mundy, Sargent, Storm, Testi, \& Wilner}]{Perez:2012ii}
P{\'e}rez, L.~M., Carpenter, J.~M., Chandler, C.~J., {et~al.} 2012, ApJL, 760,
  L17

\bibitem[{{Pfalzner} {et~al.}(2014){Pfalzner}, {Steinhausen}, \&
  {Menten}}]{2014ApJ...793L..34P}
{Pfalzner}, S., {Steinhausen}, M., \& {Menten}, K. 2014, \apjl, 793, L34

\bibitem[{Pi{\'e}tu {et~al.}(2005)Pi{\'e}tu, Guilloteau, \&
  Dutrey}]{Pietu:2005hq}
Pi{\'e}tu, V., Guilloteau, S., \& Dutrey, A. 2005, A{\&}A, 443, 945

\bibitem[{Pinilla {et~al.}(2012)Pinilla, Benisty, \&
  Birnstiel}]{Pinilla:2012ke}
Pinilla, P., Benisty, M., \& Birnstiel, T. 2012, A{\&}A, 545, A81

\bibitem[{Quanz {et~al.}(2013{\natexlab{a}})Quanz, Amara, Meyer, Kenworthy,
  Kasper, \& Girard}]{Quanz:2013ii}
Quanz, S.~P., Amara, A., Meyer, M.~R., {et~al.} 2013{\natexlab{a}}, ApJL, 766,
  L1

\bibitem[{Quanz {et~al.}(2013{\natexlab{b}})Quanz, Avenhaus, Buenzli, Garufi,
  Schmid, \& Wolf}]{Quanz:2013di}
Quanz, S.~P., Avenhaus, H., Buenzli, E., {et~al.} 2013{\natexlab{b}}, ApJL,
  766, L2

\bibitem[{Raettig {et~al.}(2015)Raettig, Klahr, \& Lyra}]{Raettig:2015vc}
Raettig, N., Klahr, H., \& Lyra, W. 2015, ApJ, in press, arxiv:1501.05364v1

\bibitem[{Raettig {et~al.}(2013)Raettig, Lyra, \& Klahr}]{Raettig:2013cn}
Raettig, N., Lyra, W., \& Klahr, H. 2013, ApJ, 765, 115

\bibitem[{Rapson {et~al.}(2015)Rapson, Kastner, Andrews, Hines, Macintosh,
  Millar-Blanchaer, \& Tamura}]{Rapson:2015km}
Rapson, V.~A., Kastner, J.~H., Andrews, S.~M., {et~al.} 2015, ApJL, 803, L10

\bibitem[{Reg{\'a}ly {et~al.}(2012)Reg{\'a}ly, Juh{\'a}sz, S{\'a}ndor, \&
  Dullemond}]{Regaly:2012fl}
Reg{\'a}ly, Z., Juh{\'a}sz, A., S{\'a}ndor, Z., \& Dullemond, C.~P. 2012,
  MNRAS, 419, 1701

\bibitem[{Ricci {et~al.}(2010{\natexlab{a}})Ricci, Testi, Natta, \&
  Brooks}]{Ricci:2010bn}
Ricci, L., Testi, L., Natta, A., \& Brooks, K.~J. 2010{\natexlab{a}}, A{\&}A,
  521, 66

\bibitem[{Ricci {et~al.}(2010{\natexlab{b}})Ricci, Testi, Natta, Neri, Cabrit,
  \& Herczeg}]{Ricci:2010gc}
Ricci, L., Testi, L., Natta, A., {et~al.} 2010{\natexlab{b}}, A{\&}A, 512, 15

\bibitem[{Rice {et~al.}(2006)Rice, Armitage, Wood, \& Lodato}]{Rice:2006ho}
Rice, W. K.~M., Armitage, P.~J., Wood, K., \& Lodato, G. 2006, MNRAS, 373, 1619

\bibitem[{{Robitaille} {et~al.}(2007){Robitaille}, {Whitney}, {Indebetouw}, \&
  {Wood}}]{2007ApJS..169..328R}
{Robitaille}, T.~P., {Whitney}, B.~A., {Indebetouw}, R., \& {Wood}, K. 2007,
  \apjs, 169, 328

\bibitem[{Robitaille {et~al.}(2006)Robitaille, Whitney, Indebetouw, Wood, \&
  Denzmore}]{Robitaille:2006cb}
Robitaille, T.~P., Whitney, B.~A., Indebetouw, R., Wood, K., \& Denzmore, P.
  2006, ApJS, 167, 256

\bibitem[{Rodmann {et~al.}(2006)Rodmann, Henning, Chandler, Mundy, \&
  Wilner}]{Rodmann:2006kk}
Rodmann, J., Henning, T., Chandler, C.~J., Mundy, L.~G., \& Wilner, D.~J. 2006,
  A{\&}A, 446, 211

\bibitem[{Roell {et~al.}(2012)Roell, Neuh{\"a}user, Seifahrt, \&
  Mugrauer}]{Roell:2012gp}
Roell, T., Neuh{\"a}user, R., Seifahrt, A., \& Mugrauer, M. 2012, A{\&}A, 542,
  A92

\bibitem[{Ros \& Johansen(2013)}]{Ros:2013ey}
Ros, K., \& Johansen, A. 2013, A{\&}A, 552, 137

\bibitem[{Rosenfeld {et~al.}(2013)Rosenfeld, Andrews, Wilner, Kastner, \&
  McClure}]{Rosenfeld:2013hv}
Rosenfeld, K.~A., Andrews, S.~M., Wilner, D.~J., Kastner, J.~H., \& McClure,
  M.~K. 2013, ApJ, 775, 136

\bibitem[{{Rucinski}(1985)}]{1985AJ.....90.2321R}
{Rucinski}, S.~M. 1985, \aj, 90, 2321

\bibitem[{Safronov(1969)}]{Safronov:1969tr}
Safronov, V.~S. 1969, Evolution of the protoplanetary cloud and formation of
  the earth and planets. English translation (1972)

\bibitem[{{Sargent} \& {Beckwith}(1987)}]{1987ApJ...323..294S}
{Sargent}, A.~I., \& {Beckwith}, S. 1987, \apj, 323, 294

\bibitem[{Schr{\"a}pler \& Henning(2004)}]{Schrapler:2004jj}
Schr{\"a}pler, R., \& Henning, T. 2004, ApJ, 614, 960

\bibitem[{Seizinger \& Kley(2013)}]{Seizinger:2013br}
Seizinger, A., \& Kley, W. 2013, A{\&}A, 551, A65

\bibitem[{Shakura \& Sunyaev(1973)}]{Shakura:1973uy}
Shakura, N.~I., \& Sunyaev, R.~A. 1973, A{\&}A, 24, 337

\bibitem[{Sicilia-Aguilar {et~al.}(2011)Sicilia-Aguilar, Henning, Dullemond,
  Patel, Juh{\'a}sz, Bouwman, \& Sturm}]{SiciliaAguilar:2011cp}
Sicilia-Aguilar, A., Henning, T., Dullemond, C.~P., {et~al.} 2011, ApJ, 742, 39

\bibitem[{Sicilia-Aguilar {et~al.}(2006)Sicilia-Aguilar, Hartmann, Calvet,
  Megeath, Muzerolle, Allen, D'Alessio, Mer{\'\i}n, Stauffer, Young, \&
  Lada}]{SiciliaAguilar:2006kj}
Sicilia-Aguilar, A., Hartmann, L., Calvet, N., {et~al.} 2006, ApJ, 638, 897

\bibitem[{Sicilia-Aguilar {et~al.}(2015)Sicilia-Aguilar, Roccatagliata, Getman,
  Rivi{\`e}re-Marichalar, Birnstiel, Mer{\'\i}n, Fang, Henning, Eiroa, \&
  Currie}]{SiciliaAguilar:2015jz}
Sicilia-Aguilar, A., Roccatagliata, V., Getman, K., {et~al.} 2015, A{\&}A, 573,
  A19

\bibitem[{Sierks {et~al.}(2015)Sierks, Barbieri, Lamy, Rodrigo, Koschny,
  Rickman, Keller, Agarwal, A{\textquoteright}Hearn, Angrilli, Auger, Barucci,
  Bertaux, Bertini, Besse, Bodewits, Capanna, Cremonese, Da~Deppo, Davidsson,
  Debei, De~Cecco, Ferri, Fornasier, Fulle, Gaskell, Giacomini, Groussin,
  Gutierrez-Marques, Guti{\'e}rrez, G{\"u}ttler, Hoekzema, Hviid, Ip, Jorda,
  Knollenberg, Kovacs, Kramm, K{\"u}hrt, K{\"u}ppers, La~Forgia, Lara,
  Lazzarin, Leyrat, Lopez~Moreno, Magrin, Marchi, Marzari, Massironi, Michalik,
  Moissl, Mottola, Naletto, Oklay, Pajola, Pertile, Preusker, Sabau, Scholten,
  Snodgrass, Thomas, Tubiana, Vincent, Wenzel, Zaccariotto, \&
  P{\"a}tzold}]{Sierks:2015db}
Sierks, H., Barbieri, C., Lamy, P.~L., {et~al.} 2015, Science, 347, aaa1044

\bibitem[{Simon \& Armitage(2014)}]{Simon:2014dr}
Simon, J.~B., \& Armitage, P.~J. 2014, ApJ, 784, 15

\bibitem[{Simon {et~al.}(2012)Simon, Beckwith, \& Armitage}]{Simon:2012dq}
Simon, J.~B., Beckwith, K., \& Armitage, P.~J. 2012, MNRAS, 422, 2685

\bibitem[{Sirono(2011{\natexlab{a}})}]{Sirono:2011cv}
Sirono, S.-I. 2011{\natexlab{a}}, ApJL, 733, L41

\bibitem[{Sirono(2011{\natexlab{b}})}]{Sirono:2011be}
---. 2011{\natexlab{b}}, ApJ, 735, 131

\bibitem[{Skrutskie {et~al.}(1990)Skrutskie, Dutkevitch, Strom, Edwards, Strom,
  \& Shure}]{Skrutskie:1990gw}
Skrutskie, M.~F., Dutkevitch, D., Strom, S.~E., {et~al.} 1990, AJ, 99, 1187

\bibitem[{{Spezzi} {et~al.}(2012){Spezzi}, {de Marchi}, {Panagia},
  {Sicilia-Aguilar}, \& {Ercolano}}]{2012MNRAS.421...78S}
{Spezzi}, L., {de Marchi}, G., {Panagia}, N., {Sicilia-Aguilar}, A., \&
  {Ercolano}, B. 2012, \mnras, 421, 78

\bibitem[{{St{\"o}rzer} \& {Hollenbach}(1999)}]{1999ApJ...515..669S}
{St{\"o}rzer}, H., \& {Hollenbach}, D. 1999, \apj, 515, 669

\bibitem[{Strom {et~al.}(1989)Strom, Strom, Edwards, Cabrit, \&
  Skrutskie}]{Strom:1989ig}
Strom, K.~M., Strom, S.~E., Edwards, S., Cabrit, S., \& Skrutskie, M.~F. 1989,
  AJ, 97, 1451

\bibitem[{Suyama {et~al.}(2008)Suyama, Wada, \& Tanaka}]{Suyama:2008bx}
Suyama, T., Wada, K., \& Tanaka, H. 2008, ApJ, 684, 1310

\bibitem[{Takeuchi \& Lin(2002)}]{Takeuchi:2002jf}
Takeuchi, T., \& Lin, D. N.~C. 2002, ApJ, 581, 1344

\bibitem[{Tamayo {et~al.}(2015)Tamayo, Triaud, Menou, \& Rein}]{Tamayo:2015dt}
Tamayo, D., Triaud, A. H. M.~J., Menou, K., \& Rein, H. 2015, ApJ, 805, 100

\bibitem[{Teiser \& Wurm(2009)}]{Teiser:2009gn}
Teiser, J., \& Wurm, G. 2009, MNRAS, 393, 1584

\bibitem[{Testi {et~al.}(2003)Testi, Natta, Shepherd, \& Wilner}]{Testi:2003dr}
Testi, L., Natta, A., Shepherd, D.~S., \& Wilner, D.~J. 2003, A{\&}A, 403, 323

\bibitem[{Testi {et~al.}(2014)Testi, Birnstiel, Ricci, Andrews, Blum,
  Carpenter, Dominik, Isella, Natta, Williams, \& Wilner}]{Testi:2014cj}
Testi, L., Birnstiel, T., Ricci, L., {et~al.} 2014, PPVI, 339

\bibitem[{Trotta {et~al.}(2013)Trotta, Testi, Natta, Isella, \&
  Ricci}]{Trotta:2013cj}
Trotta, F., Testi, L., Natta, A., Isella, A., \& Ricci, L. 2013, A{\&}A, 558,
  A64

\bibitem[{{Tsukagoshi} {et~al.}(2014){Tsukagoshi}, {Momose}, {Hashimoto},
  {Kudo}, {Andrews}, {Saito}, {Kitamura}, {Ohashi}, {Wilner}, {Kawabe}, {Abe},
  {Akiyama}, {Brandner}, {Brandt}, {Carson}, {Currie}, {Egner}, {Goto},
  {Grady}, {Guyon}, {Hayano}, {Hayashi}, {Hayashi}, {Henning}, {Hodapp},
  {Ishii}, {Iye}, {Janson}, {Kandori}, {Knapp}, {Kusakabe}, {Kuzuhara}, {Kwon},
  {McElwain}, {Matsuo}, {Mayama}, {Miyama}, {Morino}, {Moro-Mart{\'{\i}}n},
  {Nishimura}, {Pyo}, {Serabyn}, {Suenaga}, {Suto}, {Suzuki}, {Takahashi},
  {Takami}, {Takami}, {Takato}, {Terada}, {Thalmann}, {Tomono}, {Turner},
  {Usuda}, {Watanabe}, {Wisniewski}, {Yamada}, \&
  {Tamura}}]{2014ApJ...783...90T}
{Tsukagoshi}, T., {Momose}, M., {Hashimoto}, J., {et~al.} 2014, \apj, 783, 90

\bibitem[{Turner {et~al.}(2014)Turner, Fromang, Gammie, Klahr, Lesur, Wardle,
  \& Bai}]{Turner:2014ee}
Turner, N.~J., Fromang, S., Gammie, C., {et~al.} 2014, PPVI, 411

\bibitem[{van Boekel {et~al.}(2005)van Boekel, Min, Waters, de~Koter, Dominik,
  van~den Ancker, \& Bouwman}]{vanBoekel:2005fh}
van Boekel, R., Min, M., Waters, L. B. F.~M., {et~al.} 2005, A{\&}A, 437, 189

\bibitem[{van~der Marel {et~al.}(2013)van~der Marel, van Dishoeck, Bruderer,
  Birnstiel, Pinilla, Dullemond, van Kempen, Schmalzl, Brown, Herczeg, Mathews,
  \& Geers}]{vanderMarel:2013ky}
van~der Marel, N., van Dishoeck, E.~F., Bruderer, S., {et~al.} 2013, Science,
  340, 1199

\bibitem[{V{\"o}lk {et~al.}(1980)V{\"o}lk, Jones, Morfill, \&
  R{\"o}ser}]{Volk:1980vu}
V{\"o}lk, H., Jones, F.~C., Morfill, G.~E., \& R{\"o}ser, S. 1980, A{\&}A, 85,
  316

\bibitem[{Wada {et~al.}(2008)Wada, Tanaka, Suyama, Kimura, \&
  Yamamoto}]{Wada:2008eh}
Wada, K., Tanaka, H., Suyama, T., Kimura, H., \& Yamamoto, T. 2008, ApJ, 677,
  1296

\bibitem[{Wada {et~al.}(2011)Wada, Tanaka, Suyama, Kimura, \&
  Yamamoto}]{Wada:2011gd}
---. 2011, ApJ, 737, 36

\bibitem[{{Waelkens} {et~al.}(1996){Waelkens}, {Waters}, {de Graauw}, {Huygen},
  {Malfait}, {Plets}, {Vandenbussche}, {Beintema}, {Boxhoorn}, {Habing},
  {Heras}, {Kester}, {Lahuis}, {Morris}, {Roelfsema}, {Salama}, {Siebenmorgen},
  {Trams}, {van der Bliek}, {Valentijn}, \& {Wesselius}}]{1996A&A...315L.245W}
{Waelkens}, C., {Waters}, L.~B.~F.~M., {de Graauw}, M.~S., {et~al.} 1996, \aap,
  315, L245

\bibitem[{Wahlberg~Jansson \& Johansen(2014)}]{WahlbergJansson:2014ku}
Wahlberg~Jansson, K., \& Johansen, A. 2014, A{\&}A, 570, A47

\bibitem[{Weidenschilling(1977{\natexlab{a}})}]{Weidenschilling:1977p865}
Weidenschilling, S.~J. 1977{\natexlab{a}}, MNRAS, 180, 57

\bibitem[{Weidenschilling(1977{\natexlab{b}})}]{Weidenschilling:1977kq}
---. 1977{\natexlab{b}}, Ap{\&}SS, 51, 153

\bibitem[{Weidenschilling(1997)}]{Weidenschilling:1997im}
---. 1997, Icarus, 127, 290

\bibitem[{{Weintraub} {et~al.}(1987){Weintraub}, {Zuckerman}, \&
  {Masson}}]{1987ApJ...320..336W}
{Weintraub}, D.~A., {Zuckerman}, B., \& {Masson}, C.~R. 1987, \apj, 320, 336

\bibitem[{Whipple(1972)}]{Whipple:1972vv}
Whipple, F.~L. 1972, From Plasma to Planet, 211

\bibitem[{{Whitney} {et~al.}(2013){Whitney}, {Robitaille}, {Bjorkman}, {Dong},
  {Wolff}, {Wood}, \& {Honor}}]{2013ApJS..207...30W}
{Whitney}, B.~A., {Robitaille}, T.~P., {Bjorkman}, J.~E., {et~al.} 2013, \apjs,
  207, 30

\bibitem[{{Whitney} {et~al.}(2003){Whitney}, {Wood}, {Bjorkman}, \&
  {Cohen}}]{2003ApJ...598.1079W}
{Whitney}, B.~A., {Wood}, K., {Bjorkman}, J.~E., \& {Cohen}, M. 2003, \apj,
  598, 1079

\bibitem[{Williams \& Cieza(2011)}]{Williams:2011js}
Williams, J.~P., \& Cieza, L.~A. 2011, ARA{\&}A, 49, 67

\bibitem[{Windmark {et~al.}(2012{\natexlab{a}})Windmark, Birnstiel,
  G{\"u}ttler, Blum, Dullemond, \& Henning}]{Windmark:2012gi}
Windmark, F., Birnstiel, T., G{\"u}ttler, C., {et~al.} 2012{\natexlab{a}},
  A{\&}A, 540, A73

\bibitem[{Windmark {et~al.}(2012{\natexlab{b}})Windmark, Birnstiel, Ormel, \&
  Dullemond}]{Windmark:2012bg}
Windmark, F., Birnstiel, T., Ormel, C.~W., \& Dullemond, C.~P.
  2012{\natexlab{b}}, A{\&}A, 544, L16

\bibitem[{Wurm \& Blum(1998)}]{Wurm:1998kg}
Wurm, G., \& Blum, J. 1998, Icarus, 132, 125

\bibitem[{Wurm {et~al.}(2005)Wurm, Paraskov, \& Krauss}]{Wurm:2005kv}
Wurm, G., Paraskov, G., \& Krauss, O. 2005, Icarus, 178, 253

\bibitem[{Yang \& Johansen(2014)}]{Yang:2014ix}
Yang, C.-C., \& Johansen, A. 2014, ApJ, 792, 86

\bibitem[{{Yasui} {et~al.}(2010){Yasui}, {Kobayashi}, {Tokunaga}, {Saito}, \&
  {Tokoku}}]{2010ApJ...723L.113Y}
{Yasui}, C., {Kobayashi}, N., {Tokunaga}, A.~T., {Saito}, M., \& {Tokoku}, C.
  2010, \apjl, 723, L113

\bibitem[{Youdin \& Goodman(2005)}]{Youdin:2005de}
Youdin, A.~N., \& Goodman, J. 2005, ApJ, 620, 459

\bibitem[{Youdin \& Lithwick(2007)}]{Youdin:2007ef}
Youdin, A.~N., \& Lithwick, Y. 2007, Icarus, 192, 588

\bibitem[{Zhang {et~al.}(2015)Zhang, Blake, \& Bergin}]{Zhang:2015id}
Zhang, K., Blake, G.~A., \& Bergin, E.~A. 2015, ApJL, 806, L7

\bibitem[{Zhu {et~al.}(2012)Zhu, Nelson, Dong, Espaillat, \&
  Hartmann}]{Zhu:2012kr}
Zhu, Z., Nelson, R.~P., Dong, R., Espaillat, C., \& Hartmann, L. 2012, ApJ,
  755, 6

\bibitem[{Zhu {et~al.}(2011)Zhu, Nelson, Hartmann, Espaillat, \&
  Calvet}]{Zhu:2011io}
Zhu, Z., Nelson, R.~P., Hartmann, L., Espaillat, C., \& Calvet, N. 2011, ApJ,
  729, 47

\bibitem[{Zsom \& Dullemond(2008)}]{Zsom:2008je}
Zsom, A., \& Dullemond, C.~P. 2008, A{\&}A, 489, 931

\bibitem[{Zsom {et~al.}(2011)Zsom, Ormel, Dullemond, \& Henning}]{Zsom:2011fq}
Zsom, A., Ormel, C.~W., Dullemond, C.~P., \& Henning, T. 2011, A{\&}A, 534, 73

\bibitem[{Zsom {et~al.}(2010)Zsom, Ormel, G{\"u}ttler, Blum, \&
  Dullemond}]{Zsom:2010hg}
Zsom, A., Ormel, C.~W., G{\"u}ttler, C., Blum, J., \& Dullemond, C.~P. 2010,
  A{\&}A, 513, 57

\end{thebibliography}

\end{document}